\documentclass[letterpaper, aps,prd,reprint,twocolumn,superscriptaddress,nofootinbib,floatfix,longbibliography,preprintnumbers]{revtex4-2}

\usepackage{booktabs}   

\usepackage{amsmath,amssymb,mathtools,bm,braket,mathrsfs}

\usepackage{siunitx}
\DeclareSIUnit{\parsec}{pc}
\DeclareSIUnit{\decihertz}{dHz}
\DeclareSIUnit{\Msun}{\text{M}_{\odot}}

\usepackage{graphicx}
\usepackage[caption=false]{subfig}
\graphicspath{{./plot/}}

\usepackage{dcolumn} 

\usepackage[final]{microtype}
\usepackage[dvipsnames]{xcolor}
\definecolor{refblue1}{RGB}{0, 160, 200}   
\definecolor{refblue2}{RGB}{0, 190, 230}   
\newcommand{\comment}[1]{\textcolor{black}{#1}}
\usepackage[unicode]{hyperref}
\hypersetup{
  colorlinks=true,
  linkcolor=refblue1,
  citecolor=red,
  urlcolor=refblue2,
  pdfauthor={},
  pdftitle={}
}
\newcommand\inp[2]{\langle #1 \,|\, #2 \rangle}
\def \msun        {\rm{M}_\odot}
\def \mchirp      {\mathcal{M}}

\def \fmin        {f_{\rm min}}
\def \fmax        {f_{\rm max}}
\def \Mpc         {\mathrm{Mpc}}
\def \dL          {d_{\rm L}}
\def \tc          {t_{\rm c}}

\microtypesetup{protrusion=true,expansion=true}

\usepackage[capitalise,noabbrev]{cleveref}
\crefname{figure}{Fig.}{Figs.}
\Crefname{figure}{Fig.}{Figs.}

\crefformat{figure}{Fig.\thinspace(#2#1#3)}
\Crefformat{figure}{Fig.\thinspace(#2#1#3)}

\crefname{equation}{Eq.}{Eqs.}
\Crefname{equation}{Eq.}{Eqs.}

\crefformat{equation}{Eq.\thinspace(#2#1#3)}
\Crefformat{equation}{Eq.\thinspace(#2#1#3)}

\crefname{section}{Sec.}{Secs.}
\Crefname{section}{Sec.}{Secs.}

\crefformat{section}{Sec.\thinspace#2#1#3}
\Crefformat{section}{Sec.\thinspace#2#1#3}

\usepackage{xcolor}

\usepackage{orcidlink}

\usepackage[normalem]{ulem} 

\def \IITGn {Department of Physics, Indian Institute of Technology Gandhinagar, Gujarat 382055, India.\vspace*{3pt}}
\def \IUCAA {Inter-University Centre for Astronomy and Astrophysics (IUCAA), Pune 411007, India}

\begin{document}

\title{\textcolor{black}{Probing Compact Binary Coalescences in the Decihertz GW Band with IndIGO-D}}

\author{{Abhishek~Sharma}\orcidlink{0009-0007-2194-8633}}
\email{sharma.abhishek@iitgn.ac.in}
\affiliation{\IITGn}

\author{{Divya Tahelyani}\orcidlink{0000-0001-6146-7550}} 
\email{divya.tahelyani@iitgn.ac.in}
\affiliation{\IITGn}

\author{{Anand~S.~Sengupta}\orcidlink{0000-0002-3212-0475}}
\email{asengupta@iitgn.ac.in}
\affiliation{\IITGn}

\author{{Sanjit~Mitra}\orcidlink{0000-0002-0800-4626}
\vspace*{5pt}}
\email{sanjit@iucaa.in}
\affiliation{\IUCAA}

\begin{abstract}
\textcolor{black}{We present an astrophysical science-case study of IndIGO-D, a decihertz gravitational-wave mission concept, focusing on compact binary sources across the \mbox{0.1--10~Hz} band bridging LISA and next-generation ground-based interferometers. 
For these science projections, we consider a specific configuration in which three spacecraft fly in formation to form an L-shaped interferometer in a heliocentric orbit. The two orthogonal arms share a common vertex, providing a space-based analogue of terrestrial Michelson detectors. 
A central element of our study is a detailed calculation of the time-dependent antenna response of this L-shaped geometry, including the modulation produced by the orbital motion of the constellation. Assuming 1000 km arm lengths, we treat the strain-sensitivity curve supplied by the IndIGO-D collaboration as an input and investigate the astrophysical science enabled by this configuration. For the assumed sensitivity, we estimate a BNS detection rate of $\sim 76$--$767\,{\rm yr}^{-1}$.}
%
We compute horizon distances for different source classes and find that intermediate-mass black-hole binaries with masses ${10^{2}}$--$10^{3} \msun$ are detectable throughout their formation history, complementing the reach of LISA and terrestrial detectors. Binary neutron star systems are observable to a horizon \textcolor{black}{redshift} of $z \lesssim 0.3$, allowing continuous multi-band coverage with Voyager-class interferometers from the decihertz regime to merger. A Bayesian parameter-estimation study of a GW170817-like binary shows that the sky localization area improves from ${\sim 21\,\rm{deg}^2}$ at one month, \textcolor{black}{$\sim1\,\rm{deg^2}$ at one week}, down to $0.3\,\rm{deg}^2$ at six hours pre-merger.
These sky areas are readily tiled by wide-field time-domain telescopes such as the Rubin Observatory, whose $9.6\,\rm{deg}^2$ field of view and r-band depth enable high-cadence, repeated coverage of GW170817-like kilonovae at this distance and beyond. IndIGO-D exploits the rapid evolution of binaries in the decihertz band to bridge the gap between millihertz and terrestrial observations, enabling early warnings on timescales from months to hours and enhancing the prospects for multi-band and multi-messenger discoveries.
\end{abstract}

\maketitle

\section{Introduction}
\label{sec:intro}

The field of gravitational-wave (GW) astronomy has undergone a remarkable transformation since the first direct detection of astrophysical GW signals by the twin Laser Interferometer GW Observatories (LIGO) in 2015~\cite{LIGOScientific:2016aoc, LIGOScientific:2016vlm}. Over the course of its successive observing runs, the global network of ground-based detectors comprising the LIGO observatories in the United States, Virgo in Italy~\cite{VIRGO:2014yos}, and KAGRA in Japan~\cite{Akutsu2021_KAGRA_overview} have detected $\sim$300 compact binary mergers, uncovering a diverse population, including binary black holes (BBHs), binary neutron stars (BNSs), and potential neutron star black hole (NSBH) systems~\cite{Abbott_2019, Abbott_2021, gwtc_2.1, gwtc_3, LIGOScientific:2025hdt, LIGOScientific:2025slb}. These discoveries have not only confirmed the existence of stellar-mass black hole binaries but also opened new avenues for astrophysics, cosmology, and fundamental physics. For instance, the detection of GW170817~\cite{PhysRevLett.119.161101} and its associated electromagnetic counterparts~\cite{LIGOScientific:2017ync} inaugurated the era of multi-messenger astronomy with GWs, allowing for independent measurements of the Hubble constant~\cite{LIGOScientific:2017adf}, insights into the origin of heavy elements~\cite{Kasen:2017sxr}, and tests of the speed of gravity~\cite{LIGOScientific:2017zic}. 
The recently completed fourth observing run (O4) has further expanded this catalog, improving our understanding of binary population properties~\cite{LIGOScientific:2025pvj}, black hole spins, and tests of general relativity.

Despite these achievements, ground-based detectors such as LIGO, \textcolor{black}{Virgo and KAGRA} are fundamentally limited in sensitivity at low frequencies ($\lesssim 10$~Hz) due to seismic, Newtonian, and suspension thermal noise~\cite{LIGOScientific:2014pky}. As a result, they primarily observe the late inspiral, merger, and ringdown phases of compact binaries, typically lasting from a few seconds to less than a minute in-band. Extending GW observations to lower frequencies can unlock longer-duration signals, enabling us to study earlier inspiral phases with exquisite precision. The deci-Hertz (dHz; $\sim0.1$--$10$~Hz) frequency band occupies a crucial observational gap between the milli-Hertz (mHz) regime targeted by space-based detectors like the Laser Interferometer Space Antenna (LISA)~\cite{LISA:2017pwj} and the hecto to kHz range of ground-based interferometers. This intermediate frequency band hosts a rich variety of astrophysical sources, including the inspiral stages of stellar-mass compact binaries that eventually merge in the \textcolor{black}{sensitive band of terrestrial detectors}, binaries involving intermediate-mass black holes (IMBHs), and potentially new classes of exotic sources.

A detector sensitive in the dHz regime offers unique scientific advantages. Refs.~\cite{Mandel:2017pzd, Kuns:2019upi, Sedda:2019uro} highlight several fundamental questions that remain beyond the reach of both ground-based detectors and space-based observatories sensitive in the mHz band, yet could be effectively addressed by detectors in the dHz frequency range. These include uncovering the progenitors of Type Ia supernovae, probing dynamical signatures of the environments in which compact binaries merge, exploring the population and growth of intermediate-mass black holes, tracing the evolutionary pathways of stellar-mass compact objects, and investigating the formation of the light-seed black holes that may grow into the massive black holes observed at galactic centers.

In addition to addressing these questions, observations in the dHz band will capture a much longer stretch of the inspiral phase of the signal from stellar-mass compact binaries, modulated by the slowly varying antenna pattern. This extended, modulated inspiral improves source reconstruction (especially sky localization) for systems that eventually merge in the LIGO–Virgo–KAGRA (LVK) band.

While the European Space Agency (ESA)-led LISA mission will explore the low-frequency mHz domain, it primarily targets massive black hole binaries, extreme mass ratio inspirals, and galactic compact binaries. Although LISA will also observe stellar-mass BBHs that will later merge in the \comment{LVK} band (e.g., systems similar to GW150914~\cite{LIGOScientific:2016aoc, LIGOScientific:2016vlm}), such sources will take several years to transition from the LISA band to the \comment{LVK} band. In contrast, a detector operating in the dHz regime would track the same systems only hours to days before merger, making it far more practical for generating timely early-warning alerts and enabling coordinated observations. Thus, a dHz detector naturally complements both LISA and the ground-based network, bridging the observational gap and enhancing the scientific return of the global GW effort.


In this context, we study a space-based gravitational-wave observatory concept, tentatively called \mbox{\textbf{IndIGO-D}}, developed under the auspices of the Indian Initiative in Gravitational-Wave Observations (IndIGO) consortium~\cite{IndIGOWebsite, IndIGOD:2026}.
We focus on a specific realization of IndIGO-D, consisting of three spacecraft flying in formation in a heliocentric orbit to form an L-shaped interferometer with \SI{1000}{\kilo\meter} arms defined by the vertex–end separations and operating in the dHz band.
While other realizations, including geocentric configurations of the spacecraft, are also possible within the IndIGO-D concept, all results presented here pertain to the heliocentric L-shaped geometry. 
\comment{The aim of the present work is not to develop a mission design or assess the technological feasibility of IndIGO-D. Rather, we take this heliocentric L-shaped geometry and the assumed detector sensitivity as inputs, calculate its time- and frequency-dependent antenna response, and use it to make astrophysical projections for compact-binary sources in the dHz band. The L-shaped geometry considered here is distinct from the triangular configuration adopted by LISA~\cite{LISA:2017pwj} and leads to a correspondingly different time-dependent antenna response.}

The scientific opportunities offered by GW observations in the dHz band have motivated several detector concepts. Prominent examples include DECIGO~\cite{Kawamura:2006up, Sato:2017dkf, Kawamura:2020pcg}, TianGO~\cite{Kuns:2019upi}, Lunar Gravitational-Wave Antenna (LGWA)~\cite{LGWA:2020mma, Ajith:2024mie, Yelikar:2025jwh} and Laser Interferometer Lunar Antenna (LILA)~\cite{Jani:2025uaz, Creighton:2025kth}. DECIGO consists of four clusters of independent interferometric observatories orbiting in a heliocentric orbit. LGWA concept is based on the deployment of highly sensitive seismometers or inertial sensors on the lunar surface to measure the tiny tidal deformations induced by GWs. LILA will be an interferometric observatory on the lunar surface, leveraging the Moon's exceptionally low seismic and Newtonian noise environment.



\comment{We describe the heliocentric orbital configuration adopted in this study in \cref{sec:detector_geometry_and_configuration}.}
In \cref{sec:antenna_pattern_functions}, we compute the full frequency-dependent antenna response, including long-baseline propagation effects and Doppler modulation from the orbital motion. In \cref{subsec:Horizon_distance_estimates}, we estimate horizon distances for a broad range of compact binary sources using the  \comment{assumed} sensitivity and compare the reach of IndIGO-D with that of existing and proposed observatories. We highlight several scientific opportunities that will be opened up by IndIGO-D, namely, prospects of multiband observations of BNS systems in \cref{subsec:Multiband_BNS}; detectability of intermediate-mass BBHs in \cref{subsec:IMBH}; and studying environmental effects and orbital eccentricity in \cref{subsec:Environments_and_eccentricity}.
Finally, we evaluate its early-warning capabilities through Bayesian parameter-estimation studies of simulated signals from a GW170817-like BNS system in \cref{subsec:Early_warning_capabilities}. In \cref{sec:Conclusions}, 
we present our conclusions and discuss future directions.

\begin{figure*}[!hbt]
    \centering
    \includegraphics[width=0.95\linewidth]{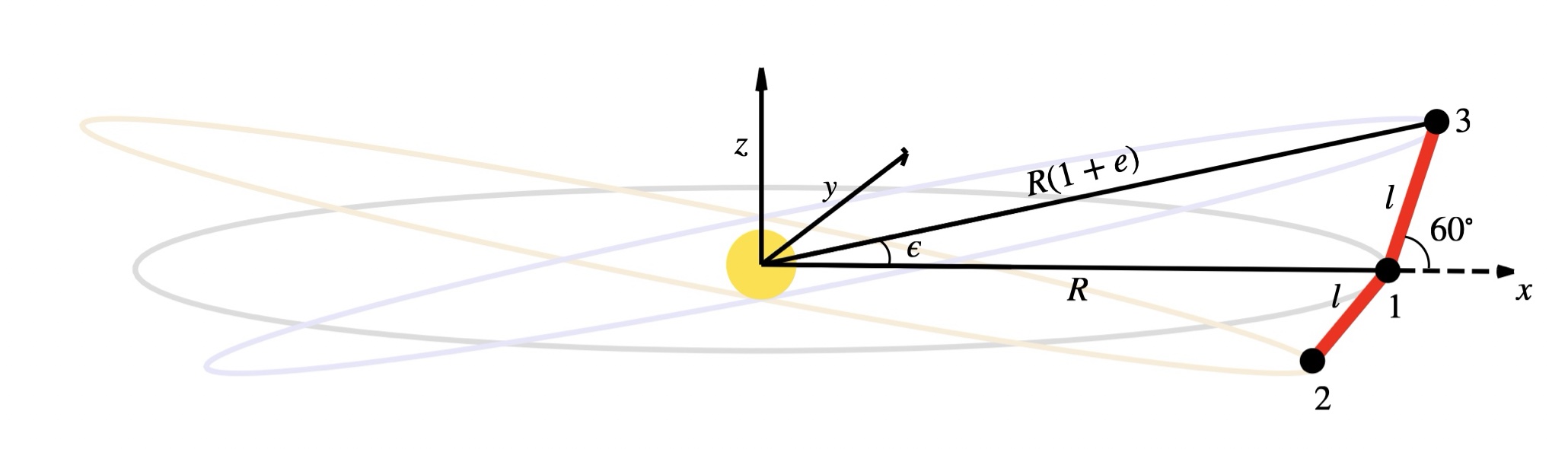}
    \caption{The orbits of the three spacecraft constituting IndIGO-D. The spacecraft 1 at the vertex orbits the Sun in the ecliptic plane at a distance of 1 AU, while the orbits of spacecraft 2 and 3 are inclined by $60^\circ$ with respect to the ecliptic. \comment{For this configuration, the inter-spacecraft separation remains nearly constant throughout the orbital motion.}}
    \label{fig:gwsat_trajectory}
\end{figure*}

\section{Detector's geometry and orbital configuration}
\label{sec:detector_geometry_and_configuration}
In the baseline configuration outlined earlier, the vertex spacecraft follows an Earth-like heliocentric orbit in the ecliptic plane, leading Earth by about \SI{20}{\degree} in the same orbital path (see~\cref{fig:gwsat_trajectory}).
The orbits of the other two spacecraft are inclined with respect to the ecliptic such that the two lines joining each spacecraft to the vertex spacecraft remain perpendicular to each other, thereby creating an L-shaped geometry.

However, it is difficult to maintain a constant inter-spacecraft separation during the orbital path of the constellation. In the context of LISA, Dhurandhar et al.~\cite{Dhurandhar:2004rv} described that the inter-spacecraft separation remains most stable along the flight when the plane of the constellation is tilted by \SI{60}{\degree} with respect to the ecliptic. \textcolor{black}{This result applies to any orbital configuration, not just the triangular configuration of LISA.} We use this result and adapt it for an L-shaped configuration to find the orbits of the spacecraft forming the IndIGO-D detector. 

The proposed orbits are such that the inter-spacecraft separation remains fixed to first order in the small dimensionless parameter $\alpha = l/2R \approx 3.34\times10^{-6}$, where $l$ is the arm-length and $R$ is the orbital radius of the vertex spacecraft.
In this study, we ignore the contribution of gravitational interactions with other heavenly objects that will affect the motion of the constellation. \comment{In comparison, for LISA, $\alpha=8.35 \times 10^{-3}$, corresponding to its $2.5\times 10^6\, \mathrm{km}$ arm-length.}

The vertex spacecraft follows a circular trajectory in the ecliptic plane ($x-y$ plane in \cref{fig:gwsat_trajectory}) and the orbit equation in terms of the coordinates $(x_1, y_1)$ in the solar system barycenter (SSB) frame are given by,
\begin{align}
\label{Eq:sc1_orbit}
x_1 = R\cos(\omega t),\qquad y_1 = R\sin(\omega t),\, \quad z_1 = 0.
\end{align}
where, $R = 1\,\rm{AU}$ denotes the orbital radius, $\omega$ denotes the orbital frequency of the vertex spacecraft and $t$ is the time. The orbits of the other two spacecraft must be eccentric and inclined by $60^\circ$ with respect to ecliptic plane to result in a fixed arm-length configuration up to first order in $\alpha$.

The orbit of the spacecraft lying at one end of the right-angled triangle (spacecraft 2) is given by,
\begin{equation}
\label{Eq:sc2_orbit}
\begin{aligned}
    x_2 &= R(\cos\psi_2 + e)\cos\epsilon, \\
    y_2 &= R\sqrt{1 - e^2}\,\sin\psi_2, \\
    z_2 &= R(\cos\psi_2 + e)\sin\epsilon,
\end{aligned}
\end{equation}
where, the eccentric anomaly $\psi_2$ satisfies, 
\begin{equation}
    \psi_2 + e\sin\psi_2 = \omega t.
    \label{Eq:anomaly_relation_sc2}
\end{equation}
The orbit of spacecraft 3 lying at the other end of the right-angled triangle is given by,
\begin{equation}
\label{Eq:sc3_orbit}
\begin{aligned}
    x_3 &= -R \sqrt{1 - e^2}\,\sin\psi_3, \\
    y_3 &=  R(\cos\psi_3 + e)\cos\epsilon, \\
    z_3 &=  R(\cos\psi_3 + e)\sin\epsilon,
\end{aligned}
\end{equation}
%
where, $\psi_3$ is given by the solution to the following equation,
\begin{equation}
    \psi_3 + e\sin\psi_3 = \omega t - \frac{\pi}{2}.
    \label{Eq:anomaly_relation_sc3}
\end{equation}
The orbital inclination $\epsilon$ can be  obtained from the geometry and is given by,
\begin{equation}
    \tan \epsilon = \frac{\frac{\sqrt{3}l}{2R}}{1 + \frac{l}{2R}} \equiv \frac{\sqrt{3}\, \alpha}{1 + \alpha},
    \label{Eq:orbital_inclination}
\end{equation}
This corresponds to the orbital inclination, $\epsilon \approx5.79\times10^{-6}$. The orbital eccentricity, $e$, can be obtained by solving $R^2(1 + e)^2 = (\sqrt{3}l/2)^2 + (R + l/2)^2$,
\begin{equation}
    e = \sqrt{1 + 2\alpha + 4 \alpha^2} -1 \approx 3.34\times 10^{-6}.
    \label{Eq:orbital_eccentricity}
\end{equation}

\begin{figure}[h]
    \centering
    \includegraphics[width=1\linewidth]{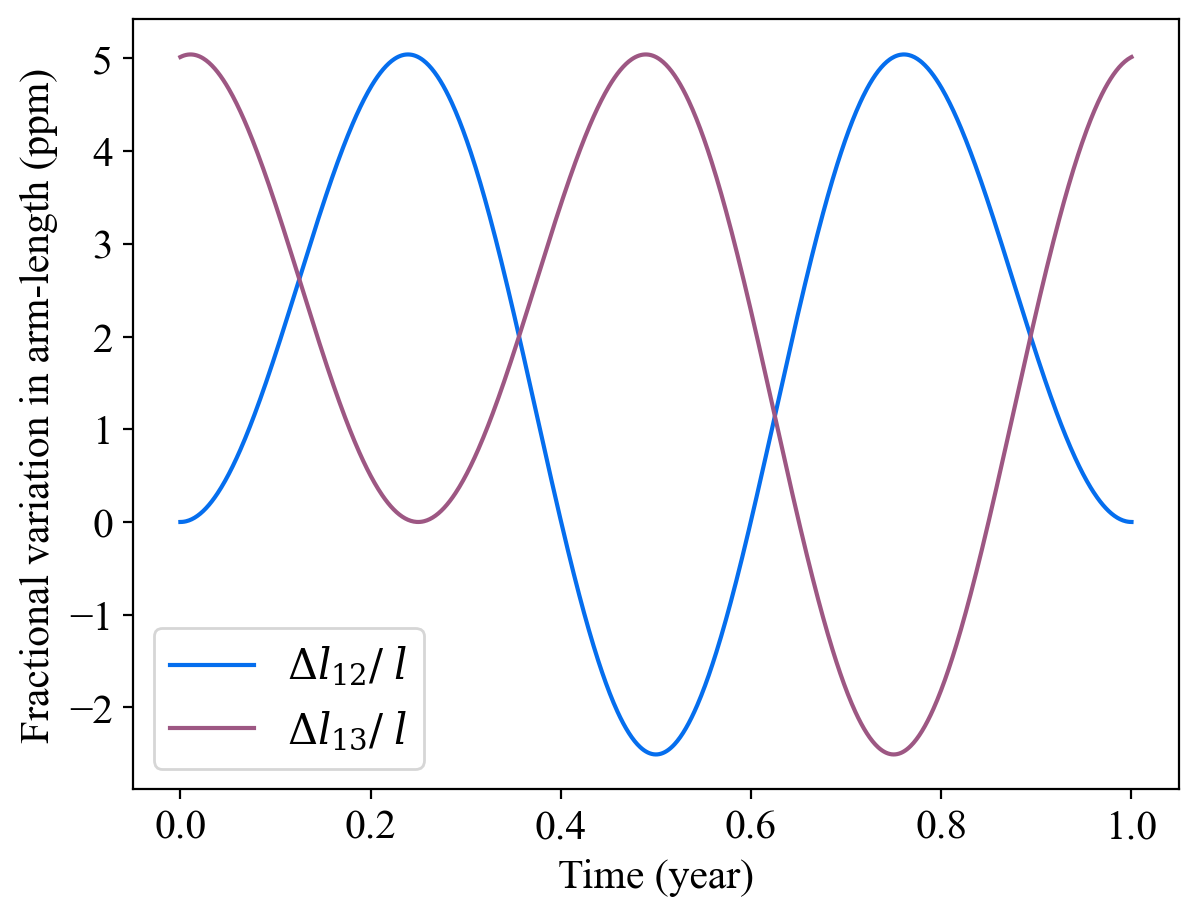}
    \caption{Flexing of the arms: fractional variation of the inter-spacecraft separation (computed using the \textcolor{black}{exact orbits}
    ) along the two arms over one orbital period of the constellation. The separation varies by ${\sim 7.5\,\mathrm{m}}$ over a drift timescale ${t_{\rm drift} \sim \mathcal{O}(\mathrm{months})}$ within a single orbital period.}
    \label{fig:breathing_modes}
\end{figure}

The time evolution of the spacecraft orbits can be computed by substituting $\psi_2, \, \psi_3,\, \epsilon$ and $e$ into Eqs.~\eqref{Eq:sc1_orbit},~\eqref{Eq:sc2_orbit}, and \eqref{Eq:sc3_orbit}.

The exact orbits of the three spacecraft result in a slight variation in the arm-length along the trajectory. \Cref{fig:breathing_modes} shows the variation in the inter-spacecraft separations in one full orbital period of the constellation.

To first order in $\alpha$,~\cref{Eq:orbital_inclination} and~\cref{Eq:orbital_eccentricity} imply that $\epsilon\approx \sqrt{3}\alpha$ and $e \approx \alpha$ and the eccentric anomalies $\psi_2$ and $\psi_3$ for the two spacecraft orbits become,
\begin{align}
\label{Eq:anomaly_relation_approximated}
\psi_2 &= \omega t - e \sin(\omega t), \nonumber \\
\psi_3 &= (\omega t - \pi/2) - e \sin(\omega t - \pi/2)\,.
\end{align}
These expressions when substituted into \cref{Eq:sc2_orbit,Eq:sc3_orbit}, result in a constant inter-spacecraft separation $l$ along the trajectory.

\begin{figure*}
    \centering
    \includegraphics[width=0.95\linewidth]{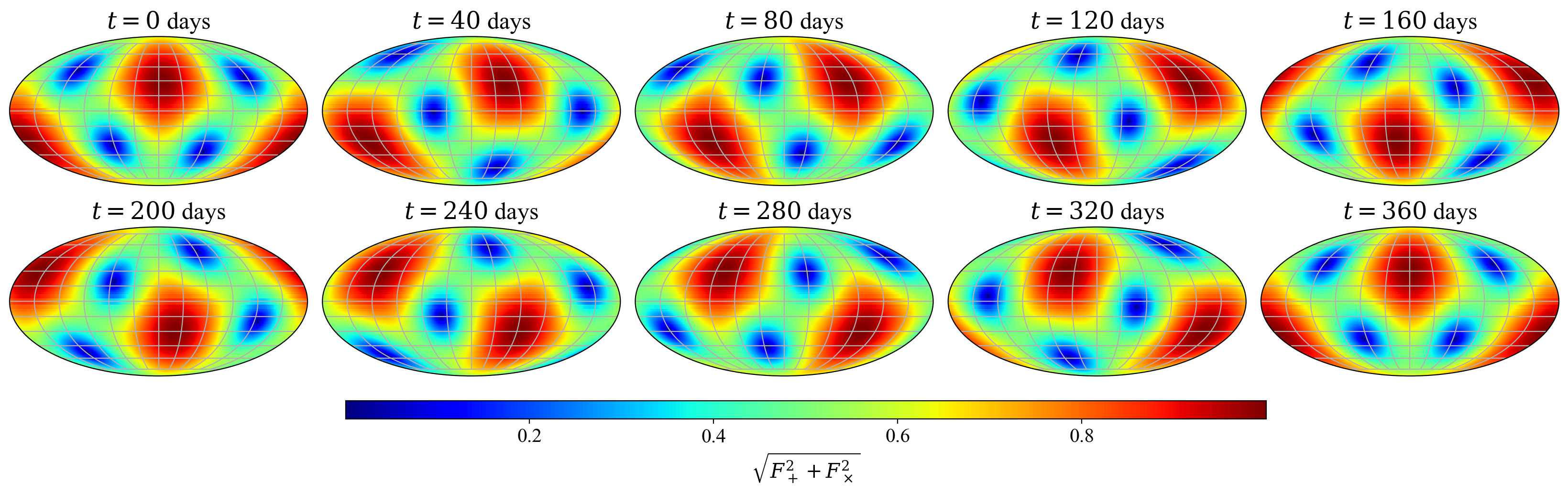}
    \caption{Variation in the magnitude of the antenna pattern functions across the whole sky in nearly one orbital period of the spacecraft constellation, assuming long-wavelength limit. Grids in the sky-map correspond to the ecliptic longitude and ecliptic latitude that marks the position in the sky in the SSB frame. Note that at any instant there exists 4 blind and 2 bright spots where the magnitude of antenna patterns are 0 and 1, respectively. The 4 blind spots lie in the detector's plane along the lines making an angle of $45^\circ$ with the detector's arms. The 2 bright spots lie perpendicular to the detector's plane, above and below the detector.}
    \label{fig:antennaPatternOverSky}
\end{figure*}

\section{Antenna Pattern Functions}
\label{sec:antenna_pattern_functions}

Antenna pattern functions describe the directional sensitivity of the detector towards the two polarizations of the GWs. Due to longer in-band duration of the GW signal, the antenna pattern will not be static. Moreover, due to large interferometric arms, the finite detector size effects will also be non-ignorable. In this section, we describe the time and frequency dependent antenna pattern functions for the envisioned IndIGO-D detector. 



For the dHz detector configuration we study here, the two interferometric baselines form
orthogonal arms of length $l_{1}$ and $l_{2}$ that meet at the vertex spacecraft. We denote by
$\hat{l}_1$ and $\hat{l}_2$ the unit vectors pointing away from this
vertex; they satisfy ${\hat{l}_1\!\cdot\!\hat{l}_2 = 0}$ and
${|\hat{l}_1| = |\hat{l}_2| = 1}$, and specify the instantaneous
orientations of the two arms. The vertex spacecraft, located at
$\vec{x}_1$ in the SSB frame, serves as the laser transmitter for the
other two spacecraft.

Using~\cref{Eq:single_arm_phase_shift}, the phase shift recorded at the output port of the Michelson interferometer is given by,
%
\begin{align}
\Delta\tilde{\phi}(f;t) 
&= \tilde{\phi}_{1}(f;t) - \tilde{\phi}_{2}(f;t) \notag \\[2pt]
&= 2\pi\, e^{-2\pi i f\,\hat{k}\cdot\vec{x}_1/c}\,
   \tilde{h}_{ij}(f)
   \Big[
      l_1^{\,} \hat l_{1}^i \hat l_{1}^j\, \mathcal{T}_1
      -
      l_2^{\,} \hat l_{2}^i \hat l_{2}^j\, \mathcal{T}_2
   \Big] \label{eq:Michelson_phase_diff}
\end{align}
where, 
$$
\mathcal{T}_1 \equiv \mathcal{T}(\hat{l}_1 \!\cdot\! \hat{k},\, f;\, t),
\qquad
\mathcal{T}_2 \equiv \mathcal{T}(\hat{l}_2 \!\cdot\! \hat{k},\, f;\, t),
$$
are the single-arm
transfer functions as shown in \cref{eq:Transfer_function}. This factor encodes the finite
light-travel time along the arm and the projection of the GW
on the arm direction. 

In Appendix~\ref{sec:appendix} we quantify the effect of the few-ppm arm-length drift that occurs over a timescale of several months. At the leading order, this drift introduces an $\mathcal{O}(10^{-6})$ fractional correction to the Michelson phase response, which is negligible. 
This level of variation is well below typical calibration uncertainties~\cite{Sun_2020} in advanced LIGO detectors and systematic errors~\cite{PhysRevD.103.124015} in state-of-the-art waveform models. 
For context, although LISA is a distinct mission with ${\sim\, 10^{9}\,\si{\meter}}$ arms and a different interferometric geometry, its calibration studies~\cite{Savalle2022Calibration} indicate strain-reconstruction uncertainties of ${\mathcal{O}\!\left(10^{-3} \text{--} 10^{-2}\right)}$, providing a useful reference scale for calibration systematics in space-based interferometers.
We therefore treat the arms as having
fixed and equal length $l$, i.e., ${l_1 = l_2 = l}$.~\footnote{In practice, a small fixed difference in the arm-length, known as Schnupp asymmetry~\cite{LIGOScientific:2014pky}, maybe maintained in the interferometer, which we are not considering here for this theoretical study.}
 
\noindent
With this simplification,~\cref{eq:Michelson_phase_diff} can now be written as,
\begin{align}
\nonumber
\Delta\tilde{\phi}(f;t) 
&= 2\pi l\, e^{-2\pi i f\,\hat{k}\cdot\vec{x}_1/c}\,
   \tilde{h}_{ij}(f)
   \Big[
      \hat l_{1}^i \hat l_{1}^j\, \mathcal{T}_1
      -
      \hat l_{2}^i \hat l_{2}^j\, \mathcal{T}_2
   \Big] \notag \\[2pt]
&= 4 \pi l\; e^{-2\pi i f{\hat{k}} \cdot \vec{x}_1/c}\;D^{ij}(\hat{l_1}, \hat{l_2}, 
    \hat{k}, f;t) \; \tilde{h}_{ij}(f) \notag \\[2pt]
&= 4 \pi l\; \tilde{h}(f;t),
\label{Eq:Michelson_phase_diff2}
\end{align}
where, $D^{ij}$ is the detector tensor: 
\begin{equation}
\label{eq:detector_tensor}
D_{ij}(\hat l_1,\hat l_2,\hat k,f;t)
:=   
    \frac12 \left (\hat l_{1}^i \hat l_{1}^j\, \mathcal{T}_1
                    -
                    \hat l_{2}^i \hat l_{2}^j\, \mathcal{T}_2 \right ),
\end{equation}
%
%
and where ${\tilde{h}(f;t) \equiv D^{ij}\tilde{h}_{ij}}$ is the effective induced strain recorded in the detector. 

For the $\SI{1000}{\kilo\meter}$ arm IndIGO-D dHz detector concept, the light-travel time 
along one arm is ${t_{\rm light} \simeq \SI{3e-3}{\second}}$. In this  
band, the GW period is on timescales of order 
$t_{\rm GW} \sim \SIrange{1}{10}{\second}$. As shown in~\cref{fig:breathing_modes}, the 
arm-length drift occurs on month-scale times, 
${t_{\rm drift} \sim 10^{6}\text{--}10^{7}\,\si{\second}}$. 
This establishes a clear 
hierarchy ${t_{\rm drift} \gg t_{\rm GW} \gg t_{\rm light},}$
under which the arm-length variations act only as a slow, multiplicative 
modulation of the Michelson response. The interferometer is effectively 
rigid on light-propagation timescales and adiabatic on the timescale of the 
GW oscillation, so the drift contributes only a negligible, slowly varying 
correction to the instantaneous response.

\textcolor{black}{We emphasize that the above argument concerns the effect of the slow orbital arm-length variation on the GW response and does not, by itself, address the coupling of laser-frequency noise caused by a residual differential arm length. The present IndIGO-D concept assumes a near-equal-arm Michelson interferometer together with active formation and interferometric control and laser-frequency stabilization. In the present study, we assume that the residual laser-frequency noise after such mitigation remains below the level represented by the adopted sensitivity curve.}

\textcolor{black}{The metre-scale variation arising from the Keplerian formation occurs only on month-long timescales and represents the slowly varying trajectory about which this active control operates. Since IndIGO-D is currently at the conceptual-design stage, a complete subsystem-level demonstration of the required arm-length and laser-frequency stability is beyond the scope of this work. A detailed implementation of the laser-noise mitigation scheme, including the possible use of time-delay interferometry (TDI)~\cite{Tinto:2020fcc} if required, will be investigated as the instrumental design is developed further.}


%
\begin{figure*}[t]
    \centering
    \includegraphics[width=0.85\linewidth]{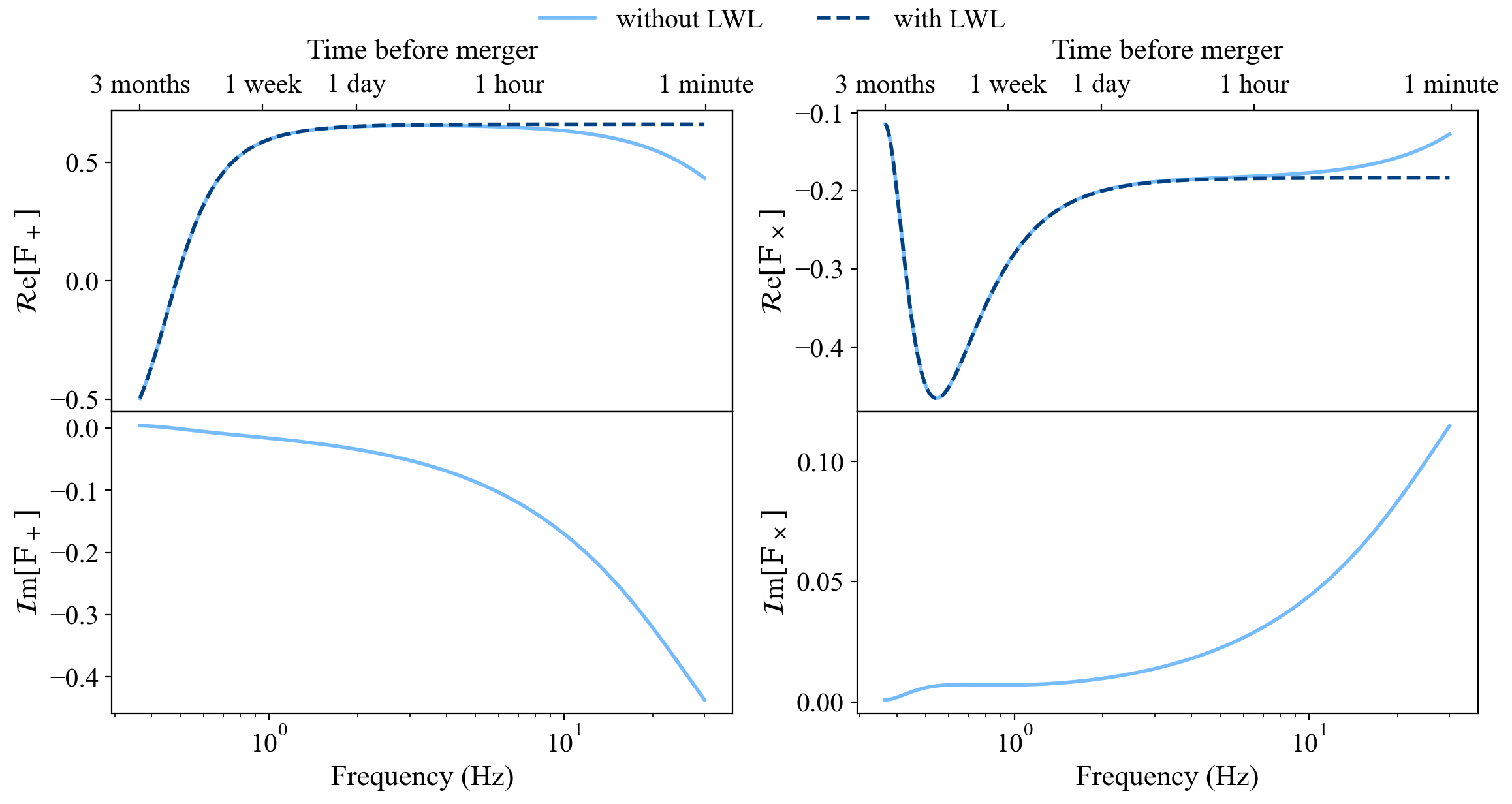}
    \caption{Comparison of the antenna pattern functions corresponding to the two polarizations, with and without assuming the LWL, evaluated in the frequency domain for a GW170817-like system as a function of frequency. Consideration of finite detector size effects results in complex-valued $F_{+,\times}$. The top and bottom panels correspond to the real and imaginary parts of $F_{+,\times}$. The modulation in the lower end of the frequency spectrum corresponds to the changing orientation and position of the detector as the binary radiates for a longer period of time in the lower frequency regime. As the inspiral rate increases, the time-dependent effects get suppressed, and the antenna pattern functions become nearly constant in LWL (dashed dark blue curve). However, consideration of finite detector size effects results in modulated $F_{+,\times}$ at the higher end of the frequency spectrum (solid light blue curve).}
    \label{fig:lwl_vs_full_response}
\end{figure*}

Expanding the metric perturbation $\tilde{h}_{ij}$ in the polarization
basis $e^{+,\times}_{ij}(\hat{k},\psi)$
(See \cref{Eq:GW_rad_frame,Eq:GW_basis_pol_tensors}),
the induced strain in the detector can be written as a linear combination
of the GW polarization amplitudes $\tilde{h}_{+,\times}(f)$:
\begin{equation}
\tilde{h}(f;t)
= e^{-2\pi i f\,\hat{k}\cdot\vec{x}_1/c}
  \sum_{a=+,\times}
  F_a(\hat{k},\psi,f;t)\,\tilde{h}_a(f).
\end{equation}
Here $F_{+,\times}$ are the antenna pattern functions, defined by the
contraction of the detector tensor with the corresponding polarization
tensors:
\begin{align}
F_+(\hat{k},\psi,f;t)
&= D^{ij}(\hat{l}_1,\hat{l}_2,\hat{k},f;t)\ e^+_{ij}(\hat{k},\psi), \\
F_\times(\hat{k},\psi,f;t)
&= D^{ij}(\hat{l}_1,\hat{l}_2,\hat{k},f;t)\ e^\times_{ij}(\hat{k},\psi).
\end{align}

\noindent
The antenna pattern functions have explicit dependence on time and frequency due to time-varying spacecraft positions and finite detector size effects, respectively. However, in the long wavelength limit (LWL), when the wavelength of the observed GW signal is very large in comparison to the length of the detector arms (${\lambda_{gw} \gg l}$, (or, equivalently the frequency $f_{gw} \ll c/2\pi l$)), the transfer functions given by \cref{eq:Transfer_function} approaches unity; and the antenna patterns become only time-dependent. 

\noindent
\Cref{fig:antennaPatternOverSky} illustrates how the combined magnitude
of the `$+$' and the `$\times$' antenna patterns varies across the sky, assuming the long-wavelength limit, over one orbital period of the
constellation.

As the source parameter estimation analysis is performed in the frequency domain due to cleaner noise modeling and more efficient likelihood evaluation, it is desirable to generate the detector frame waveforms completely in the frequency domain. The complete antenna response can be computed in the frequency domain by utilizing the time-frequency correspondence associated with the binary in a stationary phase approximation,
\begin{equation}
    t(f) = t_c - \frac{1}{2\pi}\frac{d\Psi(f)}{df},
    \label{Eq:t-fmap}
\end{equation}
where, $\Psi(f)$ is the GW phase and $t_c$ is the coalescence time. For a given set of intrinsic parameters, at any given instantaneous frequency $f$ the corresponding epoch $t$ can be evaluated using \cref{Eq:t-fmap}. Thus, the time-dependent quantities in the response function can be evaluated corresponding to a given frequency and the waveform can be computed directly in the frequency domain. 

\Cref{fig:lwl_vs_full_response} shows the comparison of the antenna response function with and without assuming LWL, evaluated entirely in the frequency domain using time-frequency mapping given by \cref{Eq:t-fmap} for a GW170817 like compact binary system. Various time stamps before the merger are also shown in the plot. The initial modulation in the low-frequency regime corresponds to the changing orientation and position of the detector. The antenna pattern factors calculated with and without assuming LWL exhibit disagreement towards the higher-end of the frequency spectrum where the finite detector size effects become important and the transfer functions (given by \cref{eq:Transfer_function}) deviate significantly from unity. Note that the antenna pattern functions \emph{without LWL} are complex valued because of the transfer functions.

\section{Science Capabilities of IndIGO-D}
\label{sec:Results}

\begin{figure*}[t]
  \centering  
  \subfloat[Noise amplitude spectral density\label{fig:noise_sensitivities}]
  {
    \includegraphics[width=0.487\textwidth]{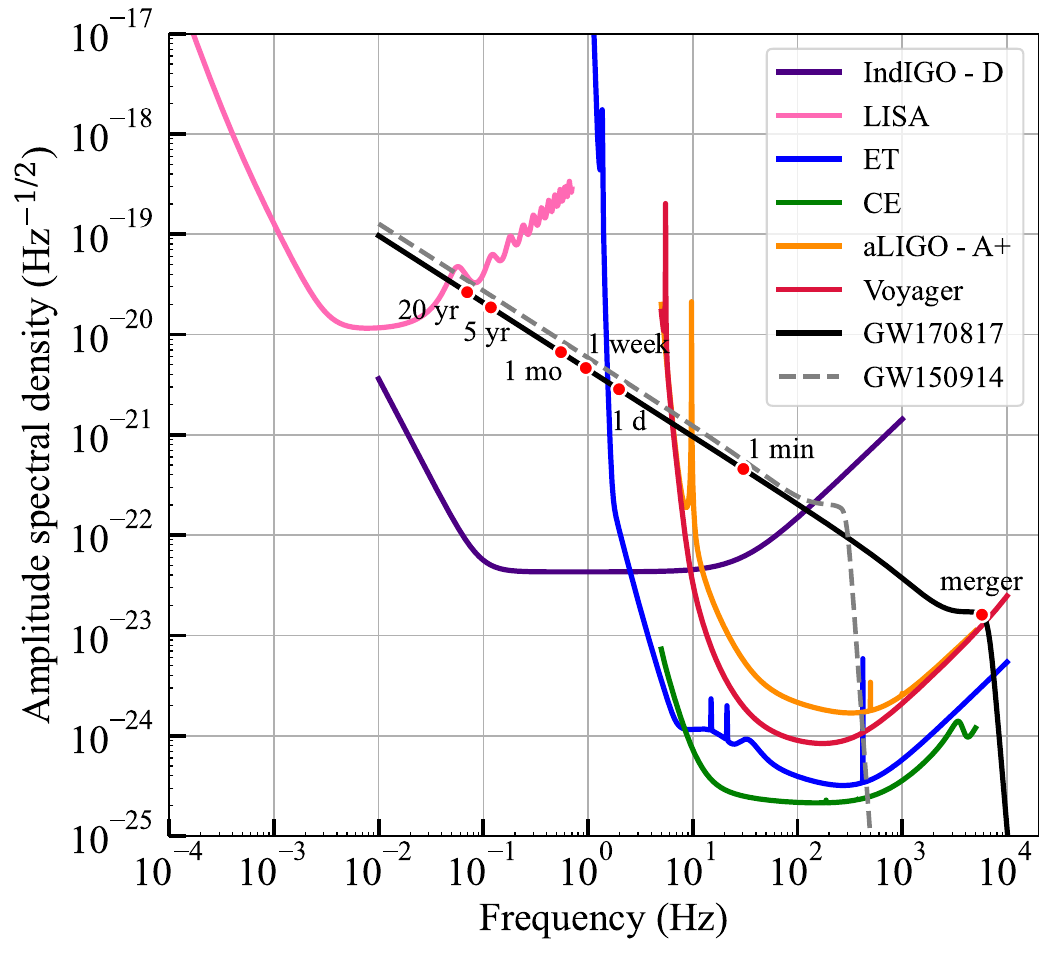}
  }
  \hfill
  \subfloat[Horizon distance\label{fig:horizon_distance}]
  {
    \includegraphics[width=0.487\textwidth]{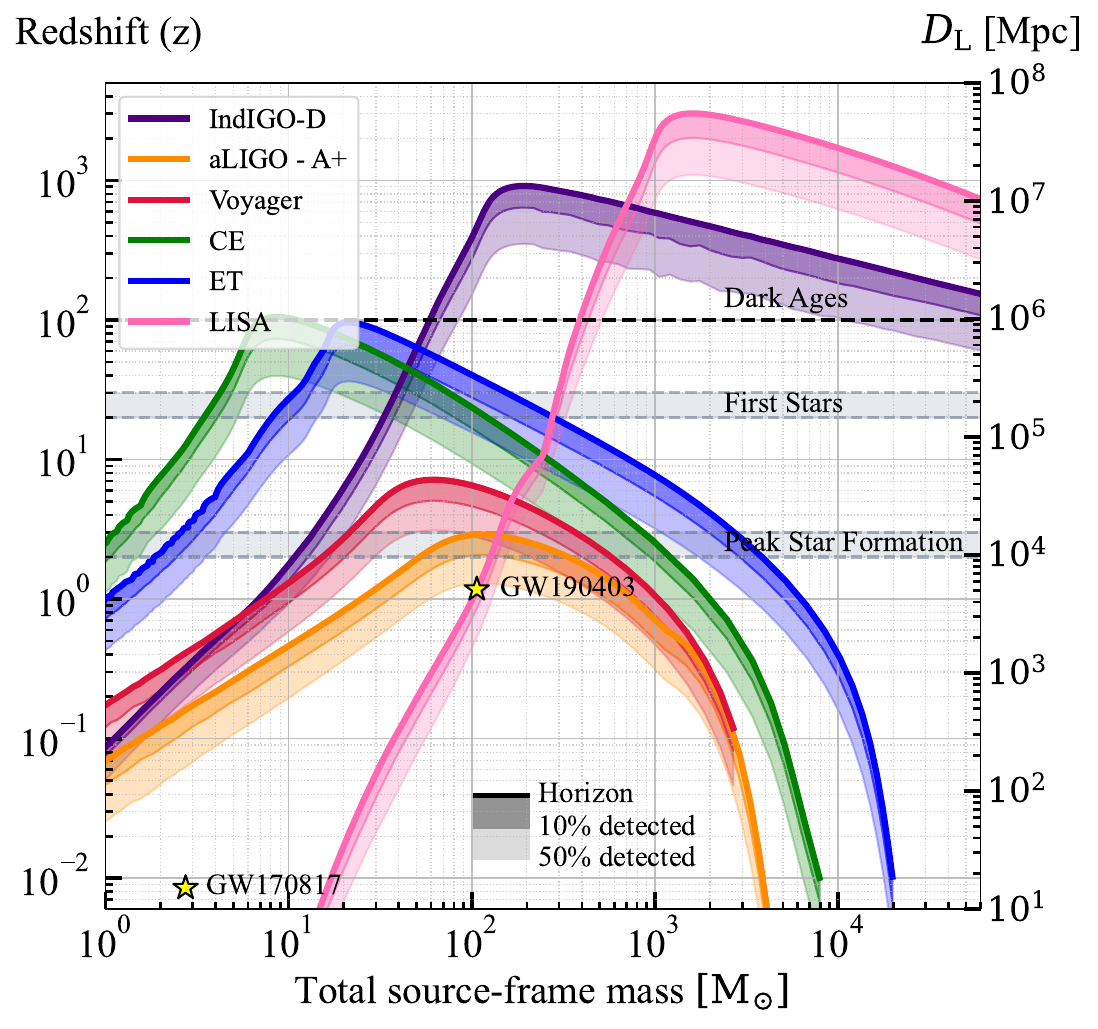}
  }
  \caption{Comparison of detector sensitivities and horizon distances. Note that IndIGO-D is sensitive to binary neutron star systems out to a horizon distance of $\sim 1000$ Mpc, similar to LIGO-Voyager~\cite{Adhikari2018Voyager}. \\
  (a) Noise sensitivities: Detector noise amplitude spectral densities (ASD) for various detectors, along with the ASD of the GW170817- and GW150914-like GW signals, given by $2\sqrt{f}\tilde h(f)$. The red-filled circles represent the time before the merger at a specific GW frequency.\\
  (b) Horizon distance (right axis) and corresponding redshift (left axis) as functions of the source-frame total mass for equal-mass, non-spinning compact binaries observed with IndIGO-D. The binaries are assumed to be face-on ($\iota=0$) and uniformly distributed over 48 sky locations. The solid curves represent the horizon distance, while the dark and light shaded regions indicate the distances within which 10\% and 50\% of the sky locations are detectable, respectively. The detection distance shown here characterizes the detector sensitivity alone and does not include assumptions about the astrophysical source population. Horizontal shaded bands mark important cosmological epochs, including the peak of cosmic star formation ($z\approx2$--$3$), the formation of the first stars ($z\approx20$--$30$), with the Dark Ages noted at higher redshifts. Star symbols indicate the masses and observed redshifts of GW170817 and GW190403, representing the nearest and most distant GW events detected to date.
  }
  \label{fig:two_panel_psd_dL}
\end{figure*}

Third-generation (3G) ground-based detectors such as Cosmic Explorer (CE)~\cite{Evans:2021gyd} and the Einstein Telescope (ET)~\cite{Punturo:2010zz} are designed to extend their sensitivity to lower frequencies than the current detector network, with CE reaching a low-frequency cutoff of approximately $5\,\mathrm{Hz}$ and ET extending down to about $2\,\mathrm{Hz}$. These upgrades are expected to extend the observable inspiral of BNS systems to tens of minutes to hours before merger. Nevertheless, even 3G instruments cannot overcome the seismic barrier below a few~\si{\hertz}, leaving a substantial region of astrophysically informative signal cycles inaccessible from the ground. The proposed IndIGO-D space-based mission would specifically target this low-frequency gap with sensitivity spanning~${\SIrange{0.1}{30}{\hertz}}$. 


\comment{For the astrophysical projections presented in this work, we adopt the
one-sided noise power spectral density (PSD) supplied by the IndIGO-D
instrumentation group~\cite{IndIGOD_repo},}
\begin{equation}
     S_n(f) = A_{\rm acc}\left(\frac{f}{f_l}\right)^{-4}
     + A_{\rm shot}\left[1+\left(\frac{f}{f_u}\right)^2\right],
     \label{eq:IndIGO_D-psd}
\end{equation}
where $A_{\rm acc}=13.10\times10^{-46}\,\si{\hertz}^{-1}$,
$A_{\rm shot}=18.47\times10^{-46}\,\si{\hertz}^{-1}$,
$f_l=0.1\,\si{\hertz}$, and $f_u=30\,\si{\hertz}$. 
\comment{The sensitivity curve adopted here assumes successful suppression of residual laser noise.}

\comment{The quantities $A_{\rm acc}$, $A_{\rm shot}$, $f_l$, and $f_u$ are
parameters of the sensitivity model supplied to us and are not derived
in the present work from an ab-initio instrumental noise budget. In
particular, we do not separately model the contributions from test-mass
acceleration noise, optical-metrology or phasemeter noise, photon shot
noise, or residual laser-frequency noise. The corresponding sensitivity
curve is shown in \cref{fig:noise_sensitivities}.}

\comment{We therefore use Eq.~(\ref{eq:IndIGO_D-psd}) as an input to the astrophysical projections presented below. Establishing the instrumental requirements necessary to realize
this sensitivity is beyond the scope of the present study.}

\Cref{fig:noise_sensitivities} shows the noise amplitude spectral densities (ASD) of various detectors, $\sqrt{S_n(f)}$, as well as the ASD of GW signals, $2\sqrt{f}\tilde h(f)$, as a function of GW frequency. The ASD of the GW170817-like signal is represented by the solid black curve, with filled red circles marking the corresponding times before merger at selected frequencies. For comparison, the gray dashed curve shows the frequency evolution of the GW150914-like BBH system across the same frequency range. Here, the waveforms are generated using the {\tt{IMRPhenomD}} waveform approximant~\cite{Khan:2015jqa}.

A GW170817-like BNS system within the IndIGO-D band would be detectable for several years prior to coalescence, enabling continuous monitoring throughout the slow early inspiral during which the majority of signal cycles accumulate. Such a system would enter the ET sensitivity band at \SI{2}{\hertz} only about a day before merger and the CE band at~\SI{5}{\hertz} about two hours prior to coalescence. Access to this prolonged low-frequency inspiral enables multi-band detection and a significant improvement in both early-warning notice and sky localization precision.

\subsection{Horizon Distances for Coalescing Compact Binary Sources}
\label{subsec:Horizon_distance_estimates}

The sensitivity of a GW detector to compact binaries is characterized by the horizon distance, defined as the farthest luminosity distance at which a source produces a signal-to-noise ratio (SNR) of~\SI{8}{}, assuming optimal orientation (face-on) and sky location. 
For a frequency-domain waveform $\tilde h(f)$, the SNR $\rho$ is given by
\begin{equation}
    \rho^2 = 4\int^{f_{\rm high}}_{f_{\rm low}} \frac{|\tilde h(f)|^2}{S_{n}(f)} \ df\,,
\end{equation}
where $S_{n}(f)$ is the one-sided noise PSD of the detector and ${f_{\rm low} \leq f \leq f_{\rm high}}$ is the detector bandwidth. For a source detected at redshift $z$, the corresponding source-frame mass is related to the detector-frame mass through ${M_{\rm source} = M_{\rm observed}/(1+z)}$. The conversion between luminosity distance and redshift depends on the assumed cosmology. In this work, we adopt the Planck 2018~\cite{Planck:2018vyg}  $\Lambda \rm CDM$ cosmology as implemented in {\tt Astropy}.

\Cref{fig:horizon_distance} illustrates, for each detector, the farthest distance and corresponding redshift at which a symmetric, non-spinning compact binary of a given total source-frame mass could be observed. To illustrate variations in detector response due to the sky location of the source, we evaluate the maximum detectable distances for 48 uniformly distributed sky positions. The upper solid line denotes the horizon distance for the most favorable sky position. The dark-shaded band encloses the luminosity distance within which \comment{$10\%$} of these sky position sources produce an SNR above the nominal threshold of 8, while the lighter-shaded region extends to the distance covering \comment{$50\%$} of detections. In this calculation, we consider a maximum in-band duration of 5 years. \comment{Horizon distance for the 3G detectors has also been discussed in Ref.~\cite{Hall:2019xmm}.}


For a GW170817-like BNS system, Advanced LIGO with A+ sensitivity~\cite{Shoemaker:2019bqt} would reach a horizon redshift of approximately $0.15$, which corresponds to  $\sim\!750\,\si{\mega\parsec}$. The 3G detectors would significantly extend this reach, with ET achieving ${z\sim\! 4}$ and CE reaching $z\sim\! 15$. The dHz space mission IndIGO-D attains a horizon redshift of $z\sim\! 0.3$. Within this overlapping redshift range ${z \lesssim\! 0.3}$, 
IndIGO-D would observe several years of the early inspiral 
of a BNS system, thereby providing precise constraints on source parameters before the system becomes detectable by terrestrial observatories. This enables coherent multi-band measurements of these sources when they would later enter the ET and CE sensitivity bands.


\subsection{Multiband Access to BNS Sources: IndIGO-D and 3G Detectors}
\label{subsec:Multiband_BNS}
Although the horizon redshift corresponds to an optimally located and face-on binary, real astrophysical binaries are distributed isotropically in both orientation and sky position. The inclination angle strongly influences the observed GW amplitude through the $(1+\cos^{2}\iota)$ and $\cos\iota$ dependencies of the plus and cross polarizations, which lowers the effective detection reach relative to the idealized horizon value. To quantify the impact of non-zero inclination and sky location on detectability and to assess the prospects for multi-band observations, we simulate a population of $10^4$ BNS systems with component source-frame masses of $1.4\,M_{\odot}$. The binary population is generated assuming isotropic inclination angles and a uniform distribution over the sky in celestial coordinates for LIGO-Voyager, which are subsequently transformed into the SSB frame for IndIGO-D.
The source redshifts are drawn uniformly in comoving volume up to $z = 0.33$, which corresponds to the approximate horizon for face-on systems for IndIGO-D as shown in \cref{fig:horizon_distance}. 

For each simulated binary, we compute the SNR in both LIGO-Voyager and IndIGO-D and determine the fraction of sources that satisfy ${\rho \geq 8}$. We find that 17.72\% of the sources are detectable by IndIGO-D and 26.04\% by Voyager. All sources detected by IndIGO-D are also detected by Voyager, so the multi-band observable samples coincide with the full IndIGO-D detection set. This demonstrates that, even after incorporating a realistic distribution of inclination angles and sky positions, a significant fraction of BNS systems within ${z \leq 0.33}$ remain jointly detectable in dHz and ground-based detector bands. IndIGO-D therefore provides strong multi-band capability when paired with Voyager or other future ground-based interferometers.


\subsection{Observational Reach for Intermediate-Mass Black-Hole Systems}
\label{subsec:IMBH}
Another key target population for IndIGO-D consists of IMBHs with total masses in the range $10^{2}\text{--}10^{5}\,\si{\Msun}$. For a nonspinning \comment{compact binary black hole source of total mass} $M$, the GW frequency associated with the innermost stable circular orbit (ISCO) is
\begin{equation}
f_{\rm ISCO} \approx \SI{4.4}{\hertz}
\left(\frac{M}{\SI{1e3}{\Msun}}\right)^{-1},
\end{equation}
which scales inversely with the total mass \comment{of the binary system}. This implies that binaries with total masses ${10^{2}\text{--}10^{4}\,\si{\msun}}$
radiate their late-inspiral and merger signal in the dHz band. In this mass regime, LISA observes only the slow, very early inspiral because its sensitivity declines above \mbox{$\sim\!{0.1}\,\si{\hertz},$} while terrestrial detectors such as \comment{LVK} and ET become sensitive to frequencies above \mbox{$\sim\!{10}\,\si{\hertz}$ and $\sim\!{1}\,\si{\hertz}$}, respectively. The gravity-gradient noise barrier around ${1}\,\si{\hertz}$ prevents ground-based detectors from detecting these systems at high redshifts. 
Consequently, probing the mass interval ${10^{2}\text{--}10^{3}\,\si{\msun}}$
robustly requires a detector optimized for the dHz band. IndIGO-D fills this observational gap precisely. Its sensitivity would enable the detection of mergers between two IMBHs as well as intermediate-mass-ratio inspirals (IMRIs) in which a stellar-mass compact object inspirals into an IMBH. 
 %

 
\textcolor{black}{The observational reach of IndIGO-D for these systems is illustrated in Fig.~\ref{fig:horizon_distance}. An equal-mass binary with a source-frame total mass of $10^{3}\,\mathrm{M}_\odot$ remains detectable out to a horizon redshift of approximately $z\sim600$, substantially exceeding the reach of both current and planned ground-based detectors in this mass range while complementing LISA's ability to observe the earlier inspiral at millihertz frequencies. We emphasize that the horizon distance is purely a measure of the detector's intrinsic sensitivity, determined by its noise PSD, and does not imply the existence of astrophysical sources at that redshift. The actual detection rate depends on the cosmic abundance of compact binaries, their formation channels, merger-rate evolution, and observational selection effects.
Current formation scenarios predict that IMBH originate at much lower redshifts, through the collapse of massive Population III stars, runaway stellar collisions in dense star clusters, or hierarchical mergers of smaller black holes \cite{Bromm:2003vv,Bromm:2013iya, Fragione:2021nhb, Klessen:2023qmc}. Population III stars are expected to form at redshifts $z\sim20–30$, while the cosmic star formation rate reaches its peak around $z\simeq2–3$~\cite{Madau:2014bja}. Due to the uncertainties associated with astrophysical population models, the local merger-rate density of IMBH binaries is generally expected to lie within the broad range of $10^{-3}-10\,{\rm Gpc}^{-3}\,{\rm yr}^{-1}$, typically predicted to reach their maximum at redshifts ${z\approx1-5}$ (see Ref.~\cite{Fragione:2022ams} and references therein). Consequently, IndIGO-D possesses sufficient sensitivity to detect IMBH mergers throughout the entire range of cosmic history relevant to their formation and evolution, providing valuable insight into their origin and their possible role as seeds of the supermassive black holes observed in galaxies today~\cite{Sesana:2009wg, Bellovary:2019nib}.}



\subsection{Probing Environmental Effects and Orbital Eccentricity}
\label{subsec:Environments_and_eccentricity}
Observations in the dHz band also offer a unique opportunity to probe the dark matter (DM) environments surrounding compact binaries. If a compact object resides within a dense distribution of DM, for example within a spike formed by the adiabatic growth of an IMBH~\cite{Gondolo:1999ef, Ullio:2001fb, Sadeghian:2013laa} or within an overdensity produced around a primordial BH seed~\cite{Green:2020jor, Boudaud:2021irr}, the surrounding matter can produce subtle modifications to the gravitational waveform~\cite{Barausse:2014tra, Eda:2013gg, Eda:2014kra, Macedo:2013qea}. Effects such as dynamical friction and particle accretion modify the inspiral rate and produce small phase shifts relative to the vacuum prediction~\cite{Kavanagh:2020cfn, Coogan:2021uqv, Derdzinski:2018qzv, Kocsis:2011dr}. Although each of these effects are individually weak, their influence accumulates over the many orbital cycles that a binary spends in the dHz band, making long-duration observations particularly sensitive to such perturbations. IMRIs with primary masses in the range ${m_1 \sim 10^{3}\text{--}10^{5}\,\si{\msun}}$
 and secondary masses ${m_2 \sim 1\text{--}10\,\si{\msun}}$, which emit GWs at $f_{\rm GW} \sim 0.1\!-\!10~\si{\hertz}$ during the final years of inspiral, are particularly promising targets. IndIGO-D’s sensitivity in this regime allows it to detect or place strong constraints on the presence of DM spikes surrounding IMBHs~\cite{Tahelyani:2024cvk}. A definitive measurement of such an overdensity would shed light on the distribution and nature of particle DM in the vicinity of compact objects.

{\color{black}
The same low-frequency access is also advantageous for measuring orbital
eccentricity~\cite{Sedda:2019uro}. The eccentricity distribution of merging compact binaries is
closely related to their formation history. Binaries formed through isolated
stellar evolution are expected to undergo substantial GW-driven
circularization before reaching the ground-based detector band
\cite{Bethe:1998bn,Kowalska:2010qg,Dominik:2012kk}. Consequently, only a very
small fraction of isolated binary-black-hole mergers are expected to retain
eccentricities $e\gtrsim0.01$ at a GW frequency of
$10\,\mathrm{Hz}$~\cite{Kowalska:2010qg}. In contrast, binaries assembled
dynamically in globular clusters, young star clusters, and nuclear star
clusters can be driven to much larger eccentricities through repeated
encounters, binary--single interactions, and gravitational captures
\cite{Samsing:2013kua,Bae2014,Hong2015,Rodriguez:2016kxx,Gondan2021,
OLeary2009,Samsing:2017xmd}. Semianalytic and numerical studies suggest that such eccentric mergers can constitute $\sim10\%$ of the BBH merger population in globular clusters, with roughly half of them retaining eccentricities $e\gtrsim0.1$ at a GW frequency of $10\,\mathrm{Hz}$~\cite{Samsing:2017xmd,Rodriguez2018,Zevin2021}. Even larger residual eccentricities may arise in nuclear star clusters, where repeated dynamical encounters and gravitational captures occur in the vicinity of supermassive
black holes
\cite{OLeary2009,Antonini:2012ad,DallAmico:2023neb,Gupte:2024jfe}.


The orbital eccentricity decreases as the binary evolves under
GW radiation reaction. At leading order, the coupled evolution
of the semi-major axis and eccentricity is described by the Peters equations
\cite{Peters:1964zz}. Eliminating the time dependence gives the
analytic relation
\begin{equation}
a(e)
=
c_0
\frac{e^{12/19}}{1-e^2}
\left(
1+\frac{121}{304}e^2
\right)^{870/2299},
\label{Eq:peters_relation}
\end{equation}
where
\begin{equation}
c_0
=
a_0
\frac{1-e_0^2}{e_0^{12/19}}
\left(
1+\frac{121}{304}e_0^2
\right)^{-870/2299},
\end{equation}
and $a_0$ and $e_0$ denote the initial semi-major axis and eccentricity.

To follow the eccentricity evolution as a function of the observed GW
frequency, we use the approximation for the peak emission frequency proposed
by Wen~\cite{Wen:2002km},
\begin{equation}
f_{\rm GW}(a,e)
=
\frac{\sqrt{GM}}{\pi}
\frac{(1+e)^{1.1954}}
{\left[a(1-e^2)\right]^{3/2}},
\label{eq:wen}
\end{equation}
where $M$ is the total mass of the binary. For a given set of initial
conditions $(a_0,e_0)$, Eqs.~\eqref{Eq:peters_relation}
and~\eqref{eq:wen} are solved together to obtain the eccentricity as a
function of the peak GW frequency.
\begin{figure}[t]
    \centering
    \includegraphics[width=\linewidth]{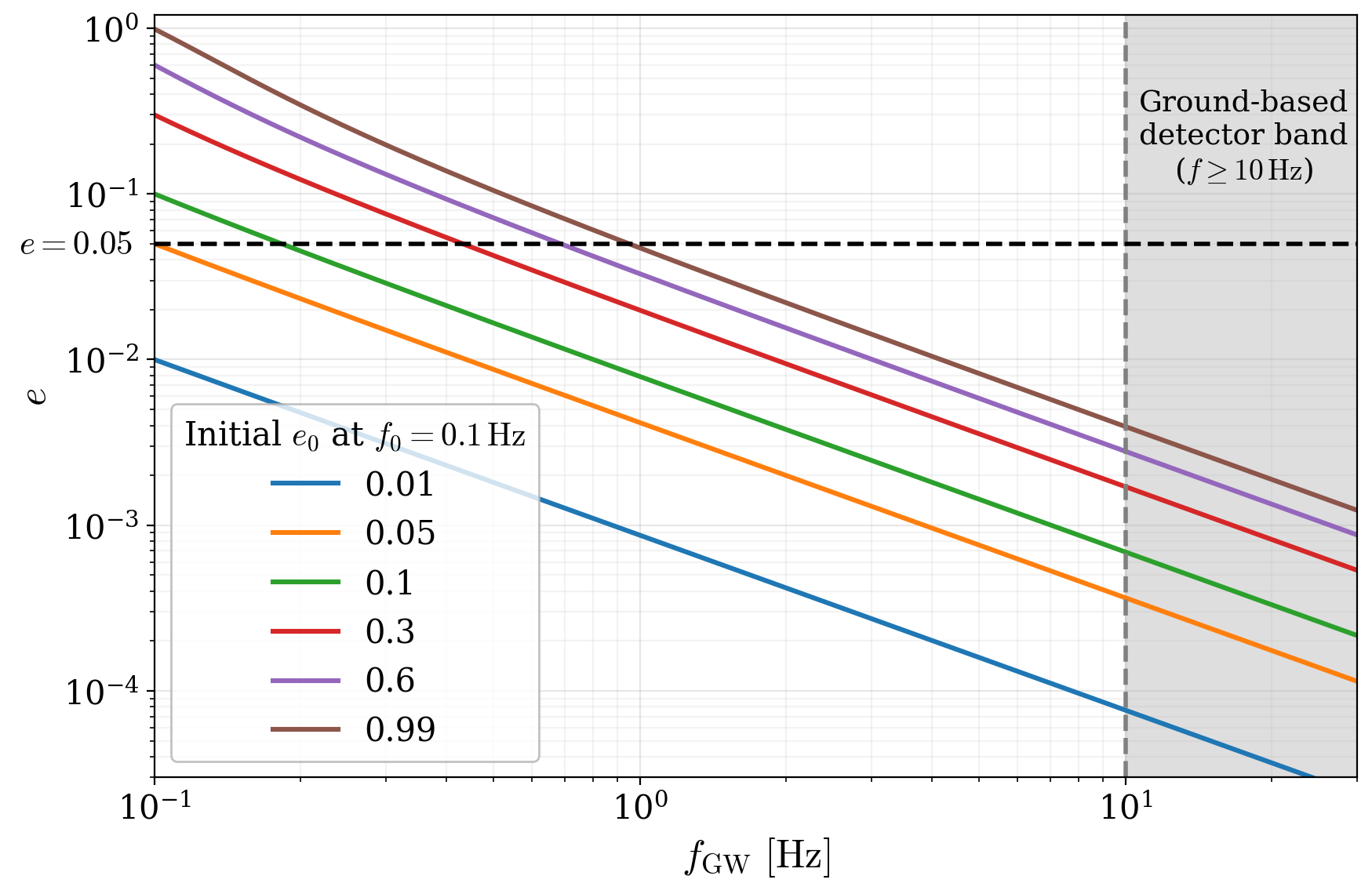}
    \caption{
    Evolution of orbital eccentricity as a function of GW frequency for different initial eccentricities $e_0$ defined at $f_0=0.1\,\mathrm{Hz}$. The grey shaded region denotes the approximate ground-based detector band, $f_{\rm GW}\geq10\,\mathrm{Hz}$. The horizontal dashed line at $e=0.05$ indicates the approximate minimum eccentricity at $10\,\mathrm{Hz}$ for which GW150914-like BBH systems were found to be distinguishable from quasi-circular binaries with Advanced LIGO and Virgo at design sensitivity
\cite{Lower:2018seu}.
    }
    \label{fig:ecc_vs_freq}
\end{figure}

Figure~\ref{fig:ecc_vs_freq} illustrates the rapid circularization of compact
binaries as they evolve toward higher frequencies. Binaries with modest
eccentricities at $0.1\,\mathrm{Hz}$ can become nearly circular by the time
they reach $10\,\mathrm{Hz}$.
Lower \textit{et al.}~\cite{Lower:2018seu} found using
Bayesian parameter estimation with the \texttt{EccentricFD} waveform~\cite{Huerta:2014eca} that
GW150914-like systems observed by Advanced LIGO and Virgo at design sensitivity can be distinguished from quasi-circular binaries for approximately $e\gtrsim0.05$ at $10\,\mathrm{Hz}$. The horizontal dashed line in Fig.~\ref{fig:ecc_vs_freq} marks this value, and the figure shows that all of the eccentricity-evolution tracks shown here fall below $e=0.05$ by the time they reach the ground-based detector band at $10\,\mathrm{Hz}$.

By observing the same systems at much lower frequencies, IndIGO-D can access their orbital evolution before GW-driven circularization erases much of this information. In addition, the long in-band duration allows eccentricity-induced phase corrections and harmonic structure to accumulate over many GW cycles, making decihertz observations a promising
probe of both compact-binary formation channels and weak environmental
effects.}
\subsection{IndIGO-D: Early warning capabilities}
\label{subsec:Early_warning_capabilities}
\begin{figure*}
    \centering
    \includegraphics[width=0.95\linewidth]{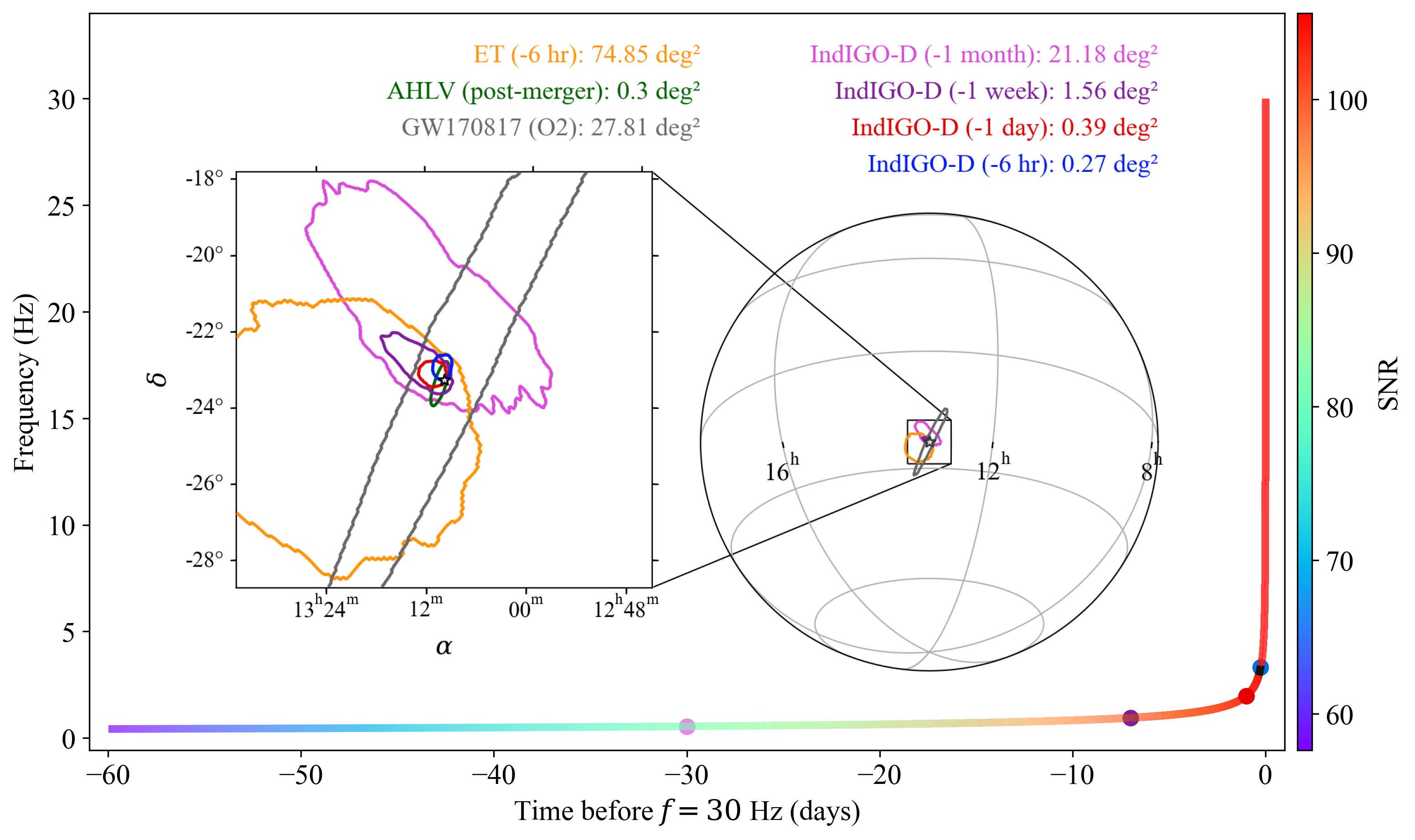}
    \caption{
    Sky-localization area (90\% credible interval) for a GW170817-like binary neutron star at different pre-merger epochs in simulated IndIGO-D data: one month, one week, one day, and six hours before merger. For comparison, we also show the localization obtained with ET at six hours before merger and the post-merger localization with the AHLV network at design sensitivity. For reference, the localization patch for the real GW170817 event observed by the HLV network during the \comment{O2} run is included. The ET sky posterior is bimodal; only the dominant mode is shown, with the secondary mode located on the opposite side of the sky. The corresponding epochs are indicated on the time–frequency track of the binary, which is colored by the accumulated signal-to-noise ratio over the three-month in-band evolution in IndIGO-D up to $30\,\si{\hertz}$.}
    \label{fig:skymaps}
\end{figure*}

Pre-merger localization of compact binary sources is crucial to not only capture the short-lived and highly beamed gamma-ray emission that decays within a few seconds right after the merger~\cite{Ziaeepour:2019mul}, but may also be important for any possible pre-merger EM emissions in binary evolution~\cite{Wang:2018xhm, Crinquand:2018tdw, Most:2020ami, Sridhar:2020uez, Most:2022ojl}.
The temporal and spectral properties of the early gamma-ray emission provide direct constraints on the formation and structure of the relativistic jet launched at merger. These observations inform models of jet energetics, angular structure, Lorentz factors, and magnetic-field configurations, thereby probing the physics of the merger remnant and jet-launching mechanisms~\cite{Mooley:2018qfh}.

A key advantage of a dHz detector is the long in-band residence time of compact binaries that will later merge in the band of terrestrial GW observatories. For BNS systems, the signal enters the IndIGO-D band several months before coalescence, allowing coherent SNR to build up well in advance of the merger. Moreover, longer in-band duration will result in Doppler modulations in the observed GW signal in IndIGO-D, which will facilitate improved sky localization of such sources.
We therefore expect IndIGO-D to provide useful pre-merger localization for nearby BNS sources, with sky areas shrinking steadily as more of the low-frequency inspiral is accumulated. 


%
To quantify the early-time localization performance with IndIGO-D, we inject a GW170817-like GW signal using {\tt{IMRPhenomD\_NRTidalv2}}~\cite{Dietrich:2018uni, Dietrich:2019kaq} waveform approximant with parameters corresponding to the maximum a posteriori (MAP) values of the posterior estimates of the real event as obtained from~\cite{Romero-Shaw:2020owr}, assuming stationary, Gaussian noise colored with the PSD. The sky position is transformed from the geocentric estimates into the SSB frame while generating the simulated data. For this analysis, we consider ${\fmin = 0.36~\si{\hertz}}$ and $\fmax = 30~\si{\hertz}$ so that the total duration of the signal is 3 months in IndIGO-D. We perform parameter estimation analyses in 13-dimensional parameter space for four distinct epochs before merger, namely, \mbox{1 month}, \mbox{1 week}, \mbox{1 day} and \mbox{6 hours} that correspond to $\fmax$ of $0.55$, $0.95$, $1.97$, and $3.32~\si{\hertz}$, respectively.
As mentioned in Appendix~\ref{sec:Bayesian_PE}, we use relative binning~\cite{Cornish:2010kf, Cornish:2021lje, Zackay:2018qdy} to accelerate the likelihood computation while sampling and use the true injection parameters as the fiducial reference waveform parameters.

\cref{fig:skymaps} shows the sky-localization area corresponding to the 90\% CI of the posterior distributions obtained for each analysis. Various epochs corresponding to distinct look-back times for which analysis is performed are also represented by dots in the time-frequency track of the binary. The time-frequency track is colored by the accumulated SNR in the IndIGO-D starting from $\fmin = 0.36~\si{\hertz}$. The values of the SNR accumulated till 1 month, 1 week, 1 day and 6 hours before merger in IndIGO-D are $82.4$, $96.8$, $102.5$ and $104.2$, respectively. 

We find that IndIGO-D can localize a binary-neutron-star source to an area of $\sim 21\,\mathrm{deg}^2$ one month before merger. This improves to $\sim 1.6\,\mathrm{deg}^2$ one week prior to merger, and shrinks further to $\sim 0.4\,\mathrm{deg}^2$ at one day and $\sim 0.3\,\mathrm{deg}^2$ at six hours before coalescence. These results are summarized in \cref{tab:sky_area}.

\begin{table}[t]
    \centering
    \renewcommand{\arraystretch}{1.15}
    \begin{tabular*}{\columnwidth}
        {@{\extracolsep{\fill}} 
        l l S[table-format=4.2] S[table-format=2.1] S[table-format=2.2]
        }
    \toprule[1pt]
        Detector & Time      & {$f_{\rm max}$} & {SNR} & {90\% sky area} \\
                 & to merger & {$\si{\hertz}$} &       & {$\rm{deg}^2$} \\
    \midrule[0.85pt]
        IndIGO-D & 1 month   & 0.55    & 82.4 & 21.18 \\
                 & 1 week    & 0.95    & 96.8 & 1.56  \\
                 & 1 day     & 1.97   & 102.5 & 0.39  \\
                 & 6 hours   & 3.32  & 104.2 & 0.27  \\
    \midrule[0.5pt]
        ET   & 6 hours      & 3.32    & 32.5 & 74.85 \\
        AHLV & post-merger  & 2048.00 & 84.2 & 0.30  \\
    \midrule[0.5pt]
        HLV (O2)  & post-merger  & 1024.00 & 32.4 & 27.81 \\
    \bottomrule[0.7pt]
    \bottomrule[0.7pt]
    \end{tabular*}
    \caption{
    Sky-localization area (90\% credible interval) for a GW170817-like binary neutron star at different pre-merger epochs, obtained with IndIGO-D and representative terrestrial high-frequency detector networks.
    For IndIGO-D, the analysis spans frequencies from ${f_{\rm low}=0.36\,\si{\hertz}}$, corresponding to an in-band duration of ${\sim 3}$ months, up to ${30\,\si{\hertz}}$.
    For ET, the analysis begins at $2\,\si{\hertz}$, while for the Advanced LIGO–Virgo (AHLV) network at design sensitivity it begins at $10\,\si{\hertz}$.
    For reference, we also show the post-merger sky-localization region of the real GW170817 event observed by the HLV network during O2.
    IndIGO-D achieves $\mathcal{O}(1)~\mathrm{deg}^2$ sky localization up to a week before merger, enabling pre-merger electromagnetic follow-up and 
    \textcolor{black}{illustrating the advantage of decihertz observations for early localization relative to the representative terrestrial configurations considered here.}
}
    \label{tab:sky_area}
\end{table}

For comparison, we compute the sky-localization area for the same source as will be observed in ET and a network of four second-generation detectors LIGO-Aundha, Hanford, Livingston, and Virgo (AHLV) at their design sensitivity. For ET, we take a lower cut-off frequency $\fmin$ of $2~\si{\hertz}$, and generate the simulated data assuming stationary, Gaussian noise colored with ET PSD, upto the epoch of 6 hours prior to merger, which corresponds to $\fmax = 3.32~\si{\hertz}$. The total in-band duration for the considered frequency interval is ${\sim 17 \; \si{hours}}$. A good sky localization is expected due to the induced modulation in the accumulated signal because of the rotation of the Earth. The total accumulated SNR is $32.5$ considering three channels of ET corresponding to three independent interferometers. It results in the localization area of $\sim$75~$\rm{deg}^2$ at the 6 hours pre-merger epoch. For AHLV network, we take $\fmin$ to be $10~\si{\hertz}$ and simulate data till the merger. It results in the localization area of $0.3$~$\rm{deg}^2$. 
%
In \cref{fig:skymaps}, we also show the sky-localization contour for GW170817 derived from publicly available posterior samples from Romero-Shaw et al.’s independent reanalysis of the GWTC-1 compact binary merger events using the Bilby Bayesian inference framework. This is not an LSC/Virgo Collaboration analysis, but an independent inference performed using open data and open-source tools~\cite{Romero-Shaw:2020owr}.

The dHz band provides substantially earlier directional information than existing or 3G high-frequency detector networks, indicating the potential of IndIGO-D for actionable pre-merger alerts and coordinated electromagnetic follow-up.

\begin{figure}[hbtp]
    \centering
    \includegraphics[width=1\linewidth]{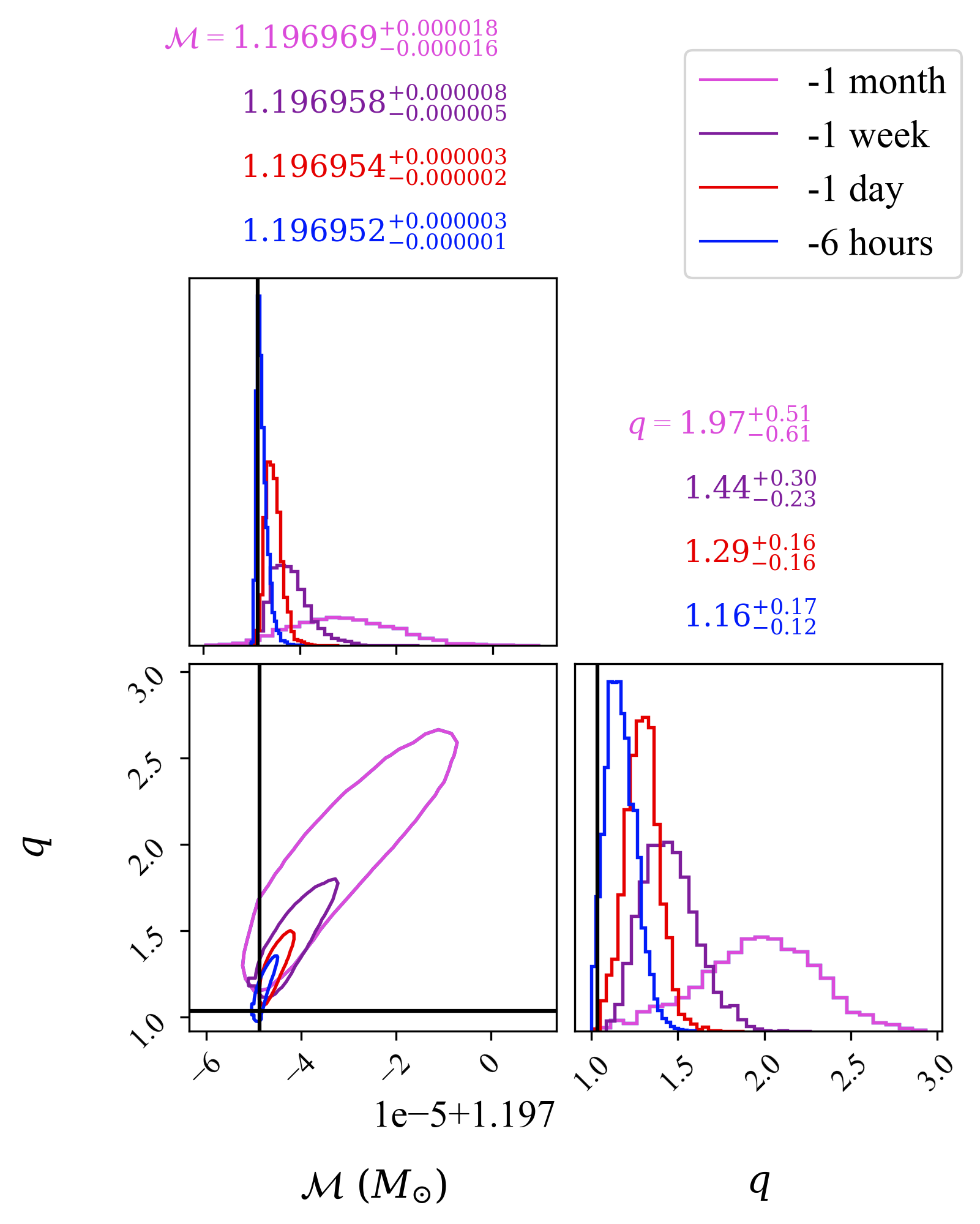}
    \caption{
    Comparison of two-dimensional marginalized posterior distributions in chirp mass and mass ratio for a GW170817-like system, obtained from simulated IndIGO-D data at different epochs before merger, as indicated in the legend. The contours show the 90\% credible regions, with colours matching those of the IndIGO-D early-warning skymaps in~\cref{fig:skymaps}.
    As expected, parameter inference becomes progressively more accurate as the signal accumulates in-band. These results can be directly compared with the corresponding sky-localization performance shown in \cref{fig:skymaps}.
    }
    \label{fig:mchirp_q_posterior}
\end{figure}

\subsubsection*{Electromagnetic follow-ups}

Studies by Nissanke et al.~\cite{Nissanke2013_endtoend} have shown that large optical telescopes with wide fields of view and deep limiting magnitudes are essential for unambiguous identification of electromagnetic counterparts to GW mergers. In our analysis, IndIGO-D can localize binary neutron star mergers to areas of \comment{$\sim 1\,\mathrm{deg}^2$} in the final few days before the signal enters the ground-based band which can be readily tiled by wide-field gamma-ray and optical telescopes. For example, the field of view (FoV) of Large Area Telescope (LAT), the primary instrument on the Fermi gamma-ray space telescope is $\sim$$7878\,\mathrm{deg}^2$~\cite{Fermi-LAT:2009ihh}. And, the Rubin Observatory/LSST, with an 8.4-m aperture and a $9.6\,\mathrm{deg}^2$ FoV per pointing, can reach a single-visit depth of $r \simeq 24.5$ for point sources~\cite{Ivezic2019_Rubin}, allowing the entire IndIGO-D localization region to be covered with only one to two pointings. For a GW170817-like kilonova with peak absolute magnitude ${M_r \simeq -16}$~\cite{Valenti2017_AT2017gfo}, this back-of-the-envelope scaling corresponds to a detection horizon of $D_L \simeq 1.26~\si{\giga\parsec}$, or ${z \approx 0.3}$ in a standard cosmology; with high-cadence revisits providing even deeper stacked limits. These numbers demonstrate that IndIGO-D’s pre-merger localization capabilities are well matched to the field of view and depth of modern wide-field optical telescopes.

\comment{In an observational regime where multiple BNS systems may be available for EM follow-up, early chirp mass measurements from IndIGO-D can provide a powerful tool for prioritizing EM follow-up. For the detector sensitivity adopted in this work, the expected detection rate can be obtained using, 
\begin{equation}
    N_{\rm{det}} = \int_0^{z_{\max}}
    R_{\rm BNS}(z)\,
    \epsilon_{\rm det}(z)
    \frac{1}{1+z}
    \frac{dV_c}{dz} dz ,
\end{equation}
where $R_{\rm BNS}(z)$ is the source-frame BNS merger-rate density, $\epsilon_{\rm det}(z)$ is the redshift-dependent detection efficiency, $dV_c/dz$ is the differential comoving volume and the factor $(1+z)^{-1}$ accounts for cosmological time dilation.}
\comment{We compute $\epsilon_{\rm det}(z)$ using a population of $5\times10^{5}$ BNS systems where the source-frame component masses are drawn uniformly over $1$--$2.5\, M_\odot$. 
The sources are distributed uniformly in comoving volume up to $z_{\max}=0.6$, and their sky locations and binary orientations are sampled isotropically.
In each redshift bin, $\epsilon_{\rm det}(z)$ is calculated as the fraction of simulated systems satisfying the detection criterion $\rho\geq8$. We model the redshift evolution of the BNS merger-rate density by convolving the Madau--Dickinson cosmic star-formation history~\cite{Madau:2014bja} with a BNS delay-time distribution following Ref.~\cite{Du:2025wto}, and normalize the resulting evolution to the median local BNS merger-rate density inferred from GWTC-5, $R_{\rm BNS}(0)=59.3^{+95.4}_{-43.9}\,{\rm Gpc}^{-3}\,{\rm yr}^{-1}$~\cite{LIGOScientific:2026ctl}. Combining the redshift-dependent merger rate with the detection efficiency obtained from our simulated BNS population, 
we estimate, for the sensitivity assumed here, an IndIGO-D detection rate of $N_{\rm det}\sim 76.4$--$767.3\,{\rm yr^{-1}}$, with a median value of $294.1\,{\rm yr^{-1}}$.
The quoted range reflects the uncertainty in the local BNS merger-rate density.
Such an event rate further motivates the use of early source characterization to identify and prioritize the most promising systems for electromagnetic follow-up.}
Systems with chirp masses indicative of long-lived remnants or enhanced mass ejection can be identified well in advance of merger and flagged as high-priority targets for intensive monitoring. This capability is particularly valuable for capturing rapidly evolving or sharply decaying EM signals.

\cref{fig:mchirp_q_posterior} shows that dHz GW observations enable an early and precise measurement of the chirp mass, with fractional uncertainties of ${\mathcal{O}(10^{-6} - 10^{-5})}$ achievable weeks to months prior to merger for a GW170817-like system. Such precision has direct implications for EM follow-up strategies. As emphasized by Margalit and Metzger~\cite{Margalit:2019dpi}, the chirp mass provides a strong discriminator of the BNS merger outcome such as prompt collapse to a black hole or the formation of a (temporary) massive neutron-star remnant, which directly affects the amount of ejecta and the resulting electromagnetic emission.

IndIGO-D could thus play a crucial role in enabling efficient and targeted multi-messenger campaigns.

\section{Conclusions and Outlook}
\label{sec:Conclusions}

In this work, we have analysed a specific heliocentric configuration of IndIGO-D, a community-led space-based interferometric gravitational-wave observatory concept developed within the IndIGO-D consortium for the decihertz band. We describe the spacecraft orbits adopted for this configuration, for which the resulting arm-length flexing is at the level of a few ppm, occurring on timescales of months.
%
This orbital configuration induces a slow cartwheel-like rotation of the detector, providing full-sky coverage over the duration in which GW signals accumulate (see~\cref{fig:antennaPatternOverSky}), along with strong Doppler and amplitude modulation of long-lived inspiral signals in the dHz band.

We describe the full antenna pattern functions of the detector.
The three-spacecraft constellation behaves effectively as a rigid interferometer on GW evolution timescales. We demonstrate that the ppm-level arm-length drifts introduce only $\mathcal{O}(10^{-6})$ corrections to the response, implying that the equal-arm approximation is valid to extremely high accuracy for the purpose of waveform modeling and parameter estimation. 

Using the \comment{assumed} sensitivity curve of IndIGO-D, we evaluate the horizon distances for a wide range of compact binary systems and compare them with the capabilities of LISA, 2G, and 3G observatories. IndIGO-D would observe the early inspirals of \comment{GW170817-like} BNS systems out to a redshift of $\sim 0.3$, which will later be observed by the LIGO Voyager or future 3G detectors such as the ET or CE, enabling multi-band observations.
\comment{For the sensitivity assumed here, and accounting for the redshift-dependent BNS merger-rate density and detection efficiency, we estimate an IndIGO-D BNS detection rate of approximately $N_{\rm det}\sim76.4$--$767.3\,{\rm yr}^{-1}$.}

At higher masses, IndIGO-D is sensitive to IMBHBs with total masses in the ${10^2 -10^4 \, \si{\msun}}$ range throughout their expected cosmic formation history for equal-mass systems. This reach complements those of both LISA and ground-based observatories and opens a new observational window onto the formation pathways of IMBHs. We also highlight the role of IndIGO-D in detecting effects of dark matter spikes around IMBHs and orbital eccentricities of compact binaries.

We quantify the early-warning capability of IndIGO-D using a detailed Bayesian inference study on simulated GW170817-like signal. We show that IndIGO-D can localize such sources within $\mathcal{O}(10)\;{\rm deg}^2$ one month, and $\mathcal{O}(1)~{\rm deg}^2$, one week before merger with accurate chirp mass measurement with uncertainties $\mathcal{O}(10^{-5})$.
The resulting sky areas are ideally suited for EM follow-up with wide-field telescopes, enabling the discovery of short GRB afterglows, kilonovae, and other transient signatures with unprecedented temporal lead time. The improvement in sky-localization will facilitate stringent measurement of the Hubble constant~\cite{Seymour:2022teq}.

Joint detections of compact binaries in the IndIGO-D and ground-based detectors also provide a powerful framework for testing general relativity~\cite{Sesana:2016ljz}. IndIGO-D would capture the early inspiral of a binary system over a large number of cycles, allowing for a precise determination of component masses and spins. Ground-based detectors would then probe the same systems during the late inspiral, merger, and ringdown, offering an independent measurement of the source’s intrinsic parameters in a different dynamical regime. Comparing the low- and high-frequency inferences would enable stringent consistency tests of GR across widely separated frequency regimes~\cite{Ghosh:2016qgn, Barausse:2016eii}.
%

By filling the spectral gap between LISA and terrestrial detectors, IndIGO-D would enable practical multiband observations in which many systems observed in the dHz regime would re-appear in the high-frequency band within days or hours, rather than after the decade-long delays characteristic of LISA-to-terrestrial detector transitions.

\appendix

\section{Effect of arm-length variation on round-trip laser phase shift}
\label{sec:appendix}

The overall antenna response of an interferometric GW detector is constructed from individual single-link Doppler measurements, which represent the fractional change in the frequency experienced by a laser beam as it propagates along the interferometer arm. We adopt the adiabatic approximation introduced in~\cite{Rubbo:2003ap}, and further used in~\cite{Cornish:2020vtw} and~\cite{Sharma:2024sfb} in the context of LISA, to model the antenna response of IndIGO-D. In this approach, the motion of the constellation is described as a sequence of stationary states. In each state, the detector is held motionless and the laser beam completes one round-trip across the detector's arms. This approximation holds when the timescale of frequency evolution of the chirping binary is longer compared to the light travel time between the arms.

The fractional change in the laser frequency induced by a GW, as it travels from the sending spacecraft $s$ located at $\vec{x}_{\text{s}}$ at barycenter time $t_\text{s}$ and arrive at a receiving spacecraft $r$ located at $\vec{x}_{\text{r}}$ at barycenter time $t_\text{r}$ is given by~\cite{Creighton:2011zz},
\begin{equation}
    y_{sr}(t) = \frac{1}{2} \frac{\hat{l}^i(t) \hat{l}^j(t)}{1 - \hat{l}(t) \cdot \hat{k}}
    \left[ h_{ij}(t_{\text{r}}, \vec{x}_{\text{r}})
    - h_{ij}(t_{\text{s}}, \vec{x}_{\text{s}}) \right],
    \label{Eq:ysr_master_relation}
\end{equation}
where, $\hat{l}(t)$ denotes the link unit vector from spacecraft $s$ to $r$ and is a time-dependent quantity due to the motion of the detector, ${h_{ij}(t, \vec{x}) \equiv h_{ij}(t - {\hat{k}} \cdot \vec{x}/c, 0)}$ is the metric perturbation describing a plane-parallel GW propagating in the $\hat k$ direction and is a linear combination of the two polarization states, $e^+_{ij}$ and $e^\times_{ij}$,
\begin{equation}
    h_{ij}(t, \vec{x}) = h_+(t, \vec{x})\; e^+_{ij} + h_{\times}(t, \vec{x})\; e^\times_{ij},
    \label{Eq:GW_rad_frame}
\end{equation}
the basis polarization states of the GW are related to the reference polarization basis by,
\begin{equation}
    \begin{pmatrix}
        e^{+}_{ij} \\
        e^{\times}_{ij}
    \end{pmatrix}
    =
    \begin{pmatrix}
        \cos 2\psi & \sin 2\psi \\
            -\sin 2\psi & \cos 2\psi
    \end{pmatrix}
    \begin{pmatrix}
        \epsilon^{+}_{ij} \\
        \epsilon^{\times}_{ij}
    \end{pmatrix},
\label{Eq:GW_basis_pol_tensors}
\end{equation}
where, $\psi$ is the polarization angle that measures the angle by which the true polarization axes of GW are rotated with respect to the reference polarization axes.

Following~\cite{Creighton:2011zz}, one can compute the fractional change in the laser frequency as it completes one round-trip across a single arm of the interferometer as a function of the GW frequency and is given by,
\begin{equation}
\tilde{\phi}(f;t) = 2\pi l \: \hat{l}^i \hat{l}^j \; e^{-2\pi i f{\hat{k}} \cdot \vec{x}_s/c}\;\tilde{h}_{ij}(f) \mathcal{T}(\hat{l} \cdot \hat{k}, f; t),
\label{Eq:single_arm_phase_shift}
\end{equation}
where,
\begin{align}
\nonumber
\mathcal{T}(\hat{l} \cdot \hat{k}, f;t) 
&:= \frac{1}{2}  \bigg\{ e^{-i \pi f l (3 + \hat{k} \cdot \hat{l})/c} \text{sinc}\left[ \pi f l (1 - \hat{k} \cdot \hat{l})/c \right] \\
&\qquad + e^{-i \pi f l (1 + \hat{k} \cdot \hat{l})/c} \text{sinc}\left[ \pi f l (1 + \hat{k} \cdot \hat{l})/c \right] \bigg\},
\label{eq:Transfer_function}
\end{align}
is the transfer function. In~\cref{Eq:single_arm_phase_shift,eq:Transfer_function}, the time dependence in the link unit vector, $\hat{l}$ and arm-length, $l$ is implicit and is not shown for brevity. In the frequency regime where $fl/c \ll 1$, the transfer function becomes a constant (unity) and this limit is regarded as the long wavelength limit (LWL).
\noindent
As evident from~\cref{fig:breathing_modes}, there is a slight variation (a few ppm) in the lengths of the two interferometric arms during the motion of the constellation over a timescale of months. We would like to quantify the effect of this drift on the measured phase difference of the laser light in one round-trip across a single arm as given by~\cref{Eq:single_arm_phase_shift}. Since the arm-length variation is small, we can write,
\begin{equation}
    l(t) = l_{0}\left[1 + \epsilon\,(t)\right],
\quad {\rm{with}} \ l_{0} = 1000\; \si{\kilo\meter},\:|\epsilon| \sim \mathcal{O}{(10^{-6})}.
\end{equation}
We can Taylor expand \(\tilde{\phi}(f,l;t)\) about \( l=l_{0} \) to get,
\begin{equation}
    \tilde{\phi}(f,l;t)
    = \tilde{\phi}_{0}(f,l_{0};t)
      + \epsilon\, l_{0} \left.\frac{\partial}{\partial l}\tilde{\phi}(f,l;t)\right|_{l=l_{0}}
      + \mathcal{O}(\epsilon^{2}),
\end{equation}
where, $\tilde{\phi}_{0}(f,l_{0};t)$ is the change in the phase of laser across one round-trip assuming static arm of length $l_0$. Therefore, the fractional departure in the phase shift due to varying arm-length from the static case is,
\begin{widetext}
    \begin{align}
    \nonumber
    \frac{\delta \tilde{\phi}(f,l;t)}{\tilde{\phi}_{0}(f,l_{0};t)} 
    &= \epsilon\, \frac{l_{0}}{\tilde{\phi}_{0}}
    \left.\frac{\partial}{\partial l}\tilde{\phi}(f,l;t)\right|_{l=l_{0}}, \\
    \nonumber
    &= \epsilon\, \frac{l_{0}}{\tilde{\phi}_{0}} \;2\pi f\, e^{-2\pi if\,\hat{k}\cdot\vec{x}_{s}/c}\,\hat{l}^{i} \hat{l}^{j} \,
        \tilde{h}_{ij}(t)\!
        \left[\left.\mathcal{T}(\hat{l} \cdot \hat{k},f,l_{0};t) + l_{0}\,\frac{\partial}{\partial l}\mathcal{T}(\hat{l} \cdot \hat{k},f,l;t)\right|_{l=l_{0}}\right], \\
    &= \epsilon \left[
    1 + \frac{x_{0}}{\tilde{\phi}_{0}(f,l_{0};t)}
    \, 2\pi f\, e^{-2\pi i f\, \hat{k}\cdot\vec{x}_{i}/c}\, \hat{l}^{i} \hat{l}^{j}\,
     \tilde{h}_{ij}(t)\,
     \left.\frac{\partial \mathcal{T}}{\partial x}\right|_{x=x_{0}}
    \right],
    \label{Eq:fractional_correction}
    \end{align}
\end{widetext}
where, ${x = \pi f l/c}$. Note that the second term in the above expression does not even contribute for low frequencies where the transfer functions remain constant. The second term starts to contribute towards the higher end of the frequency spectrum, where the LWL approximation breaks down. Upon taking the derivative of the transfer function with respect to $x$ and substituting in~\cref{Eq:fractional_correction}, one can easily check that the dominant order contributions of the second term in~\cref{Eq:fractional_correction} are $\mathcal{O}(x_{0}^2)$ and $\mathcal{O}(x_{0})$ corresponding to the real and imaginary parts, respectively. 

\noindent
In the frequency range considered for this estimate, 
i.e., ${[0.01, 10]\, \si{\hertz}}$, we have ${x_0 \lesssim 0.1}$. Thus, the fractional change is 
\begin{equation}
    \frac{\delta \tilde{\phi}(f,l;t)}{\tilde{\phi}_{0}(f,l_{0};t)} \approx \epsilon = \mathcal{O}(10^{-6}).
    \label{Eq:fractional_correction2}
\end{equation}
This implies that 
the contribution of the slowly drifting arms to the GW-induced phase response 
is suppressed by $\mathcal{O}(10^{-6})$ and are thus ignorable. This justifies the assumption of taking constant arm-lengths in~\cref{Eq:Michelson_phase_diff2}.


\section{Bayesian Parameter Estimation}
\label{sec:Bayesian_PE}

Given the strain data \( d(t) \) recorded by a detector network, containing a signal \( h(\vec{\Lambda}_\mathrm{true}; t)\) from a compact binary coalescence and additive noise \( n(t) \), the aim of parameter estimation is to determine the posterior distribution \( p(\vec{\Lambda} \mid d) \) for the source parameters. The posterior is related to the likelihood \( p(d \mid \vec{\Lambda}) \) and the prior \( p(\vec{\Lambda}) \) via Bayes’ theorem:
\begin{equation}
    p(\vec{\Lambda} \mid d) = \frac{p(d \mid \vec\Lambda)\, p(\vec \Lambda)}{p(d)},
    \label{eq:bayes_theorem}
\end{equation}
where \( p(d) \) is the evidence, which acts as an overall normalization constant. For aligned-spin BNS systems, the parameter space is typically 13-dimensional, making direct evaluation of the posterior intractable. Stochastic sampling algorithms such as Markov chain Monte Carlo (MCMC)~\cite{Foreman_Mackey_2013} and nested sampling~\cite{skilling2006nested} are therefore employed to explore \( p(\vec{\Lambda} \mid d) \). The binary parameters include a set of intrinsic source properties that govern the dynamics and phase evolution of the waveform (such as the component masses, spins and the tidal deformabilities) and a set of extrinsic parameters, namely, the source’s sky location, polarization angle \( \psi \), inclination angle \( \iota \), coalescence phase \( \varphi \), coalescence time \( \tc \) in some preferred reference frame, and luminosity distance \( \dL \). These extrinsic parameters affect the amplitude, arrival time, and modulation of the signal in the detectors.

For GW data $d^{(k)}(t) = h^{(k)}(\vec{\Lambda}_\mathrm{true}; t) + n^{(k)}(t)$ recorded at the $k^\text{th}$ detector, under the assumption of Gaussian and stationary noise $n^{(k)}(t)$, the phase-marginalized log-likelihood ratio of the signal hypothesis to the null hypothesis is given by
\begin{equation}
    \ln \mathcal{L}(\vec \Lambda) = \ln I_0\left| \sum_{k=1}^{N_{\text{d}}} \langle h^{(k)}(\vec \Lambda) \mid\boldsymbol{d}^{(k)}\rangle \right| - \frac{1}{2} \sum_{k=1}^{N_{\text{d}}} \, \left\Vert h^{(k)}(\vec \Lambda)\right\Vert^2,
    \label{eq:logl_basic}
\end{equation}
where, $I_0[\cdot]$ is the zeroth-order modified Bessel function of the first kind, $N_d$ is the number of detectors in the network and $\inp{a}{b}$ is the noise-weighted inner product of two time domain signal `vectors' $a(t)$ and $b(t)$, 
\begin{equation}
    \displaystyle \inp{a}{b} = 4 \:\Re\int_{\fmin}^{\fmax} \frac{\tilde{a}^{\ast}(f) \: \tilde{b}(f)}{S_n(f)}\: df,
    \label{eq:innerProduct}
\end{equation}
where, $\fmin$ and $\fmax$ are the lower and upper cut-off frequencies; ${\left\Vert a \right\Vert^2 = \inp{a}{a}}$ denotes the norm 
and $\tilde{a}(f)$ denotes the Fourier transform of $a(t)$. 
In order to accelerate the sampling process, we adopt the heterodyning likelihood~\cite{Cornish:2010kf,Cornish:2021lje} or the relative binning method~\cite{Zackay:2018qdy} as implemented in the {\tt{PyCBC Inference}}~\cite{Biwer:2018osg} library. 
We use the {\tt{dynesty}}~\cite{Speagle:2019ivv} nested sampling package as implemented in {\tt{Bilby}}~\cite{Ashton:2018jfp} with {\tt{acceptance-walk}} sampling method, {\tt{nlive}} = 3000 and {\tt{naccept}} = 60. The prior distributions for various parameters and their ranges used to obtain the results presented in \cref{subsec:Early_warning_capabilities} for various analyses with IndIGO-D are summarized in \cref{tab:prior_IndIGO-D_analysis}. Note that the sky location is measured in the SSB frame in terms of ecliptic longitude and ecliptic latitude while sampling the posterior distribution, which are later transformed in the celestial coordinates.
\begin{table}[t]
    \def\arraystretch{1.15}
    \centering
    \begin{tabular}{l l l}
    \toprule[1pt]
        Parameters     &\:\:  Range    &   Prior distribution \\     
        \midrule[1pt]
        $\mathcal{M}$  &   $[\mchirp_{\rm{inj}} \pm 10^{-3}~\msun]$           &   $\propto \mathcal{M}$ \\
        $q$            &   $[1, \;q_{\rm{inj}} + 2]$            & $ \propto \left [ (1 + q)/q^3 \right ]^{2/5}$ \\
        $\chi_{1z,\; 2z}$    &   $[-0.05,\; 0.05]$    & Uniform  \\
        $\dL$          & $[1,\;1000] \, \Mpc$            & Uniform in volume\\
        $t_c$          & $[t_c^{\rm{inj}} \pm 60~\rm{s}]$ & Uniform\\
        $\lambda$       & $[0, 2\pi]$            & Uniform\\
        $\beta$       & $[-\pi/2, \pi/2]$        & $\sin^{-1} \left [ {\text{Uniform}}[-1,1]\right ]$\\
        $\iota$        & $[0, \pi]$             & Uniform in $\cos \iota$\\
        $\psi$         & $[0, 2\pi]$            & Uniform \\
        $\Lambda_{1, 2}$       & $[0,\; 5000]$            & Uniform \\
    \bottomrule[1pt]
    \bottomrule[1pt]
    \end{tabular}
    \caption{The prior distributions over various parameters and their corresponding range used in the parameter estimation with IndIGO-D. The chosen prior distributions over $\mchirp$ and $q$ correspond to uniform distributions over component masses. Priors over ecliptic longitude and latitude correspond to the isotropic direction across all sky.
    }
    \label{tab:prior_IndIGO-D_analysis}
\end{table}

\acknowledgments
\comment{We thank the anonymous referee for useful comments which have led to substantial improvements in the manuscript.} A.~Sharma, D.~Tahelyani and A.~Sengupta gratefully acknowledge the hospitality of IUCAA, Pune, during our visits under the IUCAA Associates Program, where part of this research was conducted. We are grateful to the HPC facility at IIT Gandhinagar for access to computational resources, and especially to Raviraj Sukhadiya for reliable support in maintaining the cluster. A.~Sharma acknowledges IIT Gandhinagar for the senior research fellowship. D.~Tahelyani acknowledges the support from the Inspire fellowship from the Department of Science and Technology of India, under fellowship number DST/INSPIRE Fellowship/[IF210643].
%
We thank our colleagues in the IndIGO consortium for suggestions and helpful discussions and 
\textcolor{black}{providing the IndIGO-D noise curve used in this work.}

\noindent
We acknowledge the National Supercomputing Mission (NSM) for providing computing resources of ‘PARAM Ananta’ at IIT Gandhinagar, which is implemented by C-DAC and supported by the Ministry of Electronics and Information Technology (MeitY) and Dept. of Science and Technology (DST), Govt. of India.

\bibliography{references}
\end{document}